\documentclass[twocolumn]{aastex6}
\usepackage{apjfonts}
\usepackage{aas_macros, xspace} 
\received{January 25, 2021}
\revised{March 28, 2021}
\accepted{April 10, 2021}

\newcommand{\eq}{\,=\,}
\def\ts     {\thinspace}
\def\kms    {\ts km\ts s$^{-1}$}

\def\msol   {$M_{\odot}$}
\def\lsol   {$L_{\odot}$}

\def\eco    {{\rm CO}($J$=5$\to$4)}

\def\ico    {{\rm CO}($J$=9$\to$8)}

\shorttitle{GADOT Galaxy Survey:\ Dense Gas and Feedback in $z$=2--6 Dusty Starbursts}
\shortauthors{Riechers et al.}

\begin{document}

\title{The GADOT Galaxy Survey:\ Dense Gas and Feedback in {\em Herschel}-Selected Starburst Galaxies at Redshifts 2 to 6}

\author{Dominik A.\ Riechers\altaffilmark{1}}
\author{Asantha Cooray\altaffilmark{2}}
\author{Ismael P\'erez-Fournon\altaffilmark{3,4}}
\author{Roberto Neri\altaffilmark{5}}

\altaffiltext{1}{Cornell University, Space Sciences Building, Ithaca, NY 14853, USA}
\altaffiltext{2}{Department of Physics and Astronomy, University of California, Irvine, CA 92697, USA}
\altaffiltext{3}{Instituto de Astrofisica de Canarias, E-38200 La Laguna, Tenerife, Spain}
\altaffiltext{4}{Departamento de Astrofisica, Universidad de La Laguna, E-38205 La Laguna, Tenerife, Spain}
\altaffiltext{5}{Institut de RadioAstronomie Millim\'etrique, 300 Rue
  de la Piscine, Domaine Universitaire, F-38406 Saint Martin d'H\'eres,
  France}

 \email{riechers@cornell.edu}

\begin{abstract}

We report the detection of 23 OH$^+$ 1$\to$0 absorption, emission, or
P-Cygni-shaped lines and \ico\ emission lines in 18 {\em Herschel-}
selected $z$=2--6 starburst galaxies with the Atacama Large
Millimeter/submillimeter Array (ALMA) and the NOrthern Extended
Millimeter Array (NOEMA), taken as part of the Gas And Dust Over
cosmic Time (GADOT) Galaxy Survey. We find that the \ico\ luminosity
is higher than expected based on the far-infrared luminosity when
compared to nearby star-forming galaxies. Together with the strength
of the OH$^+$ emission components, this may suggest that shock
excitation of warm, dense molecular gas is more prevalent in distant
massive dusty starbursts than in nearby star-forming galaxies on
average, perhaps due to an impact of galactic winds on the gas. OH$^+$
absorption is found to be ubiquitous in massive high-redshift
starbursts, and is detected toward 89\% of the sample. The majority
of the sample shows evidence for outflows or inflows based on the
velocity shifts of the OH$^+$ absorption/emission, with a comparable
occurrence rate of both at the resolution of our observations. A small
subsample appears to show outflow velocities in excess of their escape
velocities. Thus, starburst-driven feedback appears to be important in
the evolution of massive galaxies in their most active phases. We find
a correlation between the OH$^+$ absorption optical depth and the dust
temperature, which may suggest that warmer starbursts are more compact
and have higher cosmic ray energy densities, leading to more efficient
OH$^+$ ion production.  This is in agreement with a picture in which
these high-redshift galaxies are ``scaled-up'' versions of the most
intense nearby starbursts.

\end{abstract}


\section{Introduction} \label{sec:intro}

The physical properties and chemical composition of the interstellar
medium (ISM) set the initial conditions for star formation in galaxies
through cosmic history. Following the initial detections of molecular
gas at high redshift ($z$$>$1) three decades ago (Brown \& Vanden Bout
\citeyear{brown91}; Solomon et al.\ \citeyear{solomon92}), most
studies in the early universe to date have focused on low- to
mid-level CO rotational transitions, or the [C{\sc{II}}] fine
structure line (see Carilli \& Walter \citeyear{cw13} for a
review). These studies have recently resulted in the first
measurements of the ``cold gas history of the universe'', i.e., the
comoving volume density of molecular gas over $\sim$13\,billion years
of cosmic time (e.g., Walter et al.\ \citeyear{walter14,walter20};
Riechers et al.\ \citeyear{riechers19,riechers20b}; Decarli et
al.\ \citeyear{decarli20}; see also Klitsch et
al.\ \citeyear{klitsch19}; Lenkic et
al.\ \citeyear{lenkic20}). Molecular ions were first detected at high
redshift fifteen years ago through HCO$^+$, a tracer of the dense gas
phase (Riechers et al.\ \citeyear{riechers06a, riechers10b};
Garcia-Burillo et al.\ \citeyear{gb06}). Thanks to the emergence of
sensitive submillimeter telescopes like the Atacama Pathfinder
Experiment (APEX), the {\em Herschel Space Observatory,} and more
recently, the Atacama Large Millimeter/submillimeter Array (ALMA),
studies of the ISM in the nearby universe have undergone a revolution,
leading to the discovery of many missing links in the formation
mechanisms of some of the key species, such as OH$^+$.

The reactive molecular ion OH$^+$ is a sensitive probe of the gas
phase oxygen chemistry in the ISM and the heating and excitation
mechanisms associated with star formation and active galactic nuclei
(AGN). In particular, OH$^+$ reacts quickly with H$_2$ when present
(or with free electrons through dissociative recombination), such that
high ionization rates are required to retain a substantial abundance
in the molecular phase of the ISM. Since the first interstellar
detection of submillimeter-wave OH$^+$ absorption toward Sagittarius
B2 (Wyrowski et al.\ \citeyear{wyrowski10}), it has been found that
the absorbing gas appears to reside primarily in the atomic (H{\sc I})
gas phase (e.g., Gerin et al.\ \citeyear{gerin10}; Neufeld et
al.\ \citeyear{neufeld10}). This conclusion is supported by the
observed OH$^+$/H$_2$O$^+$ ratios, which indicate comparatively low
gas density (e.g., Benz et al.\ \citeyear{benz10}). OH$^+$ in diffuse
gas is formed primarily through ionization by cosmic rays, and thus,
is used to infer the cosmic ray ionization rate in atomic hydrogen gas
(e.g., Hollenbach et al.\ \citeyear{hollenbach12}; Indriolo et
al.\ \citeyear{indriolo15}, and references therein).

Outside the Milky Way, OH$^+$ absorption has been detected in nearby
starbursts like Arp\,220 and M82 (Rangwala et
al.\ \citeyear{rangwala11}; Kamenetzky et
al.\ \citeyear{kamenetzky12}; Gonzalez-Alfonso et
al.\ \citeyear{ga13}). More recently, these studies could be extended
to the distant universe, through the detection of OH$^+$ absorption or
P-Cygni profiles in five starburst galaxies at $z$=2.3--6.3 (Riechers
et al.\ \citeyear{riechers13b,riechers20c}; Indriolo et
al.\ \citeyear{indriolo18}; Berta et al.\ \citeyear{berta20}). Based
on the kinematic information from the line profiles referenced to the
systemic redshifts (typically measured from CO emission lines), the
absorption components in these intense starbursts have been found to
be associated with large-scale outflows or inflows, providing direct
evidence for galaxy-scale feedback. OH$^+$ has also been detected
purely in emission in the Orion Bar (van der Tak et
al.\ \citeyear{vdt13}), as well as in nearby AGN host galaxies like
Mrk\,231 and NGC\,1068 (van der Werf et al.\ \citeyear{vdw10};
Spinoglio et al.\ \citeyear{spinoglio12}). OH$^+$ emission has
recently even been detected in two AGN host galaxies at $z$=3--6 (Li
et al.\ \citeyear{li20}; Stanley et al.\ \citeyear{stanley20}).

Besides newly discovered species, the study of high-$J$ CO transitions
has also been transformed by {\em Herschel} through the wide-spread
detection of nearly complete CO line ladders up to CO($J$=13$\to$12)
and beyond in large samples of nearby galaxies (e.g., Kamenetzky et
al.\ \citeyear{kamenetzky14}; Rosenberg et
al.\ \citeyear{rosenberg15}). These investigations have lead to a
significantly more comprehensive understanding of the different
heating and cooling mechanisms in the ISM of star-forming galaxies,
and on the role of AGN vs.\ ultraviolet and mechanical heating for the
excitation of the CO line ladders. Building upon some of the initial
high-redshift detections of high-$J$ CO lines (beyond $J$=7$\to$6) in
AGN host galaxies (e.g., Downes et al.\ \citeyear{downes99};
Wei\ss\ et al.\ \citeyear{weiss07}; Bradford et
al.\ \citeyear{bradford09}), the advent of ALMA and the upgraded
NOrthern Extended Millimeter Array (NOEMA) have now made it possible
to more routinely extend the studies of broadly-sampled CO line
ladders to dusty starburst galaxies (DSFGs), i.e., the most active,
luminous massive star-forming galaxies to exist since the onset of
galaxy formation (see, e.g., Hodge \& da Cunha \citeyear{hd20} for a
recent review).

A key aspect that has only been facilitated by few studies to date is
that telescopes like ALMA and NOEMA offer sufficiently broad bandwidth
toward studies of high-redshift galaxies to allow simultaneous
coverage of high-$J$ CO lines and other key species, including OH$^+$.
Most relevant for this work, it is possible to simultaneously observe
the \ico\ and OH$^+$ 1$_1$$\to$0$_1$ and 1$_2$$\to$0$_1$ lines at
rest-frame $\sim$1\,THz over a broad range in redshift. Since \ico\ is
most likely primarily associated with the warm, dense gas in the
star-forming regions, it is a good indicator of the systemic
redshift. Thus, the kinematic information entailed in the OH$^+$ lines
can be referenced ``in-band'' to the systemic value from CO with
minimal calibration biases.

We here report the detection of 20 ground state OH$^+$ lines and 16
\ico\ lines toward a sample of bright {\em Herschel-}selected DSFGs
at $z$=2.3--5.3 observed with ALMA and NOEMA. We also include
observations of three OH$^+$ lines and two \ico\ lines in two {\em
  Herschel-} selected targets from the literature (ADFS-27 at $z$=5.66
and HFLS3 at $z$=6.34; Riechers et
al.\ \citeyear{riechers13b,riechers20c}), which were obtained as part
of our broader follow-up efforts of the parent sample from the Gas And
Dust Over cosmic Time (GADOT) Galaxy Survey (D.~Riechers et al., in
prep). We study relations between these diagnostic lines and the
physical properties of the ISM and star-forming regions in distant
galaxies in the context of our current understanding of high-mass
galaxy formation. We present the data and their calibration in Section
2, before discussing the immediate results and presenting a broader
analysis, informed by models and scaling relations, in Sections 3 and
4. A summary and conclusions are given in Section 5. We use a
concordance, flat $\Lambda$CDM cosmology throughout, with
$H_0$\eq69.6\,\kms\,Mpc$^{-1}$, $\Omega_{\rm M}$\eq0.286, and
$\Omega_{\Lambda}$\eq0.714.

\section{Data} \label{sec:data}

\subsection{Sample Properties}

The full GADOT Galaxy Survey master sample is composed of $\sim$100
bright, {\em Herschel}-selected galaxies from the {\em HerMES}+{\em
  HeLMS}+{\em HeRS} and {\em H-ATLAS} surveys (Eales et
al.\ \citeyear{eales10}; Oliver et al.\ \citeyear{oliver12}) with
spectroscopic redshifts. The majority of observations, including those
presented here, focus on the subsample from the former survey, which
itself consists of the ``RARE'' (i.e., the 500\,$\mu$m-brightest
sources in the survey area; e.g., Conley et al.\ \citeyear{conley11};
Riechers et al.\ \citeyear{riechers11d}; Wardlow et
al.\ \citeyear{wardlow13}; Bussmann et al.\ \citeyear{bussmann15};
Nayyeri et al.\ \citeyear{nayyeri16}; Gomez Guijarro et
al.\ \citeyear{gomez19}) and ``red'' (i.e., sources with
$S_{250}$$<$$S_{350}$$<$$S_{500}$; e.g., Riechers et
al.\ \citeyear{riechers13b}; Dowell et al.\ \citeyear{dowell14};
Asboth et al.\ \citeyear{asboth16}) subsamples. The ``red'' subsample
includes both strongly-lensed sources which overlap with the ``RARE''
sample at the higher fluxes, and unlensed (or at most weakly-lensed)
hyper-luminous infrared galaxies (HyLIRGs, or ``Titans''; Riechers et
al.\ \citeyear{riechers17,riechers20c}).

For our ALMA Atacama Compact Array/Morita Array (ACA) follow-up
campaign, we selected 19 of the 250--500\,$\mu$m-brightest sources
with spectroscopic redshifts in equatorial fields, with the goal to
understand their detailed interstellar medium
properties.\footnote{Southern HerMES fields were not included due to a
  lack of spectroscopic redshifts at the time of the initial study.}
Three of these sources overlap with the ``red'' source
catalog,\footnote{These sources are HeLMS-10, 28, and 65.} and
three more sources are consistent with being ``red'' within the
1$\sigma$ uncertainties of the 250--500\,$\mu$m
measurements.\footnote{These sources are Orochi and HeLMS-59 and 62.}
From this sample of 19 sources, we here selected 13 sources for which
the \ico\ and OH$^+$(1$_1$$\to$0$_1$) lines can be observed in a
common tuning in bands 5--7, where observations in ACA standalone mode
were offered by the observatory.\footnote{One additional source was
  observed in \ico, but the OH$^+$ line could not be covered by the
  tuning. We thus excluded the source from this work.}

Three additional sources were observed in the same lines as part of
our NOEMA follow-up campaign of strongly-lensed ``red'' sources. These
galaxies were selected for an in-depth follow-up campaign, because
they make up all of the currently known strongly-lensed $z$$>$5
starbursts in the HerMES fields. As such, they are a natural extension
of the ACA sample towards higher redshifts.

Spectroscopic redshifts for the full sample were obtained as part of
our broader follow-up campaign (see Table~\ref{t1} for details).  With
the exception of the weakly-lensed merger HXMM-01, all sources in this
sample are strongly gravitationally lensed, as revealed by ALMA, VLA,
and Submillimeter Array (SMA) observations at 0.05$''$--0.4$''$
spatial resolution (Bussmann et al.\ \citeyear{bussmann15};
Amvrosiadis et al.\ \citeyear{amvrosiadis18}; Geach et
al.\ \citeyear{geach18}; D.\ Riechers et al., in
preparation). Together with observations of this new sample of 16
sources, we include previous \ico\ and OH$^+$ detections in the
$z$=5.66 and 6.34 hyper-luminous dusty starbursts ADFS-27 and HFLS3
which are also part of our master sample of $\sim$100 sources
(Riechers et al.\ \citeyear{riechers13b,riechers20c}). Neither of them
is strongly lensed.

\begin{figure*}
\begin{deluxetable}{ l c c c c c c c c c c }
%\vspace{-7mm}

\tabletypesize{\scriptsize}
\tablecaption{ALMA/ACA and NOEMA Observations. \label{t1}}
\tablehead{
Name\tablenotemark{a} & HerMES/HeRS$^\star$ ID & redshift & $\nu_{\rm obs}^{\rm CO(9-8)}$ & band & $N_{\rm ant}$ & $\theta_{\rm maj}$$\times$$\theta_{\rm min}$ & observing & $t_{\rm  on}$ & complex gain & bandpass/ \\
     &           &          & (GHz)                  &      &            & 290\,$\mu$m & dates      & (min)  & calibrator & flux$^\dagger$ calib. }
\startdata
{\em ALMA/ACA (6.4\,hr):} & &        &          &   &    &        &      &    &           &  \\
HeLMS-5 (HELMS8)         & J004714.1+032453  & 2.2919\tablenotemark{b} & 314.9890 & 7 & 12 & 4\farcs8$\times$3\farcs4 & 2017 Sep 07 & 33.3 & J0108+0135 & J2253+1608 \\
                         &                   &                         &          &   &    & &             &      & & Uranus$^\dagger$ \\
HeLMS-9 (HELMS18)        & J005159.4+062240  & 2.3934\tablenotemark{c} & 305.5674 & 7 & 11 & 5\farcs4$\times$3\farcs9 & 2018 Nov 20 & 44.4 & J0108+0135 & J2258--2758 \\
                         &                   &        & 169.8202\tablenotemark{d} & 5 & 11 & 11\farcs2$\times$6\farcs3\tablenotemark{d} & 2018 Dec 04 & 17.7 & J0108+0135 & J2258--2758 \\
HeLMS-10 (HELMS10)       & J005258.6+061317  & 4.3726\tablenotemark{b} & 193.0001 & 5 & 12 & 10\farcs6$\times$5\farcs5 & 2018 Nov 27 & 27.3 & J0108+0135 & J0237+2848 \\
HeLMS-28 (HELMS4)        & J004410.2+011821  & 4.1625\tablenotemark{b} & 200.8547 & 5 & 12 & 7\farcs8$\times$5\farcs1 & 2018 Nov 27 & 49.5 & J0108+0135 & J2253+1608 \\
                         &                           &                         &          &   &    & 8\farcs2$\times$5\farcs4\tablenotemark{e} &             &      & & \\
HeLMS-44 (HELMS2)        & J233255.5--031134 & 2.6895\tablenotemark{c} & 281.0442 & 7 & 11 & 6\farcs8$\times$3\farcs6 & 2018 Oct 25 & 14.6 & J2323--0317 & J2253+1608 \\
HeLMS-45 (HELMS7)        & J232439.4--043934 & 2.4726\tablenotemark{c} & 298.5983 & 7 & 10 & 7\farcs7$\times$3\farcs2 & 2018 Oct 16 & 17.7 & J2301--0158 & J2258--2758 \\
HeLMS-59 (HELMS6)        & J233620.7--060826 & 3.4346\tablenotemark{b} & 233.8232 & 6 & 12 & 8\farcs3$\times$4\farcs2 & 2018 Oct 21 & 28.3 & J2301--0158 & J2258--2758 \\
                         &                           &                         &          &   &    & 9\farcs0$\times$4\farcs5\tablenotemark{e} &             &      & & \\
HeLMS-61 (HELMS15)       & J233255.7--053424 & 2.4022\tablenotemark{c} & 304.7770 & 7 & 11 & 5\farcs8$\times$3\farcs3 & 2018 Oct 23 & 16.2 & J2323--0317 & J2253+1608 \\
HeLMS-62 (HELMS5)        & J234051.3--041937 & 3.5027\tablenotemark{b} & 230.2868 & 6 & 12 & 8\farcs1$\times$4\farcs4 & 2018 Oct 22 & 17.7 & J0006--0623 & J2258--2758 \\
                         &                           &                         &          &   &    & 8\farcs5$\times$4\farcs7\tablenotemark{e} &             &      & & \\
HeLMS-65 (HELMS\_RED\_31)& J002737.3--020759 & 3.7966\tablenotemark{g} & 216.1765 & 6 & 12 & 9\farcs2$\times$4\farcs5 & 2018 Oct 27 & 35.3 & J2358--1020 & J2253+1608 \\
J0210+0016 (HeRS-1)      & J020941.1+001557$^\star$ & 2.5534\tablenotemark{f} & 291.8085 & 7 & 12 & 5\farcs3$\times$3\farcs5 & 2018 Oct 21 & 15.1 & J0217+0144  & J0237+2848 \\
HXMM-01 (XMM-01)         & J022016.5--060143 & 2.3079\tablenotemark{b} & 313.4655 & 7 & 10 & 5\farcs2$\times$3\farcs4 & 2018 Oct 24 & 44.4 & J0241--0815 & J0423--0120 \\
                         & &        &          &   & 12 & & 2018 Oct 28 &      & & J0522--3627 \\
Orochi (HXMM-02/XMM-06)  & J021830.6--053131 & 3.3903\tablenotemark{b} & 236.1826 & 6 & 11 & 6\farcs9$\times$4\farcs1 & 2018 Nov 01 & 23.7 & J0217--0820 & J0423--0120 \\
                         &                           &                         &          &   &    & 7\farcs3$\times$4\farcs4\tablenotemark{e} &             &      & & \\
\tableline
{\em PdBI/NOEMA (14.2\,hr):} & &        &          &   &    &          &    &    &             & \\
HeLMS-34 (HELMS\_RED\_4/ & J002220.9--015520 & 5.1614\tablenotemark{g} & 168.2917 & 2 & 5D & 5\farcs1$\times$3\farcs7 & 2015 Sep 07 & 110  & B0106+013 & 3C454.3 \\
HELMS29)                 & &        &          &   &    &         &    &     &             & LkHA101$^\dagger$ \\
HXMM-30 (XMM-30)         & J022656.6--032709 & 5.094\tablenotemark{g}  & 170.1530 & 2 & 6D & 4\farcs0$\times$2\farcs9 & 2015 Sep 21 & 270  & B0215+015 & 3C454.3 \\
                 & &        &          &   &    &         &    &     &             & LkHA101$^\dagger$ \\
                 & &        &          &   &    &         &    &     &             & MWC\,349$^\dagger$ \\
HLock-102 (Lock-102)     & J104050.5+560652  & 5.2915\tablenotemark{b} & 164.8116 & 2 & 5D & 3\farcs9$\times$3\farcs5 & 2013 Jun 12 & 470  & B0954+556 & 3C84 \\
                         & &        &          &   &  &  & 2013 Jul 04 &      & B0954+556 & 3C279/3C454.3 \\
                         & &        &          &   &  &  & 2013 Aug 17 &      & B0954+658 & LkHA101$^\dagger$ \\
                         & &        &          &   &  &  & 2015 Sep 10 &      & B0954+658 & MWC\,349$^\dagger$ \\
\tableline
{\em GADOT literature sources:} & &        &          &   &    &          &    &    &             & \\
ADFS-27                  & J043657.7--543810 & 5.6550\tablenotemark{g} & 155.8095 & 4 & 46\tablenotemark{h} & 0\farcs56$\times$0\farcs49 & 2018 Nov 03 & 76   & J0441--5154 & J0519--4546 \\
HFLS3                    & J170647.8+584623  & 6.3369\tablenotemark{b} & 141.3284 & 2 & 6S\tablenotemark{i} & 4\farcs4$\times$3\farcs1 & 2011 Dec 01/11 & 200  & B1637+574 & 3C345/3C454.3 \\
                         & &        &          &   &  &  & &      & B1849+670 & MWC\,349$^\dagger$ \\
\enddata
\tablenotetext{\rm a}{Alternative source names are from Wardlow et al.\ (\citeyear{wardlow13}), Asboth et al.\ (\citeyear{asboth16}), or Nayyeri et al.\ (\citeyear{nayyeri16}). Additional alternative source identifications are used throughout the literature; we here adopt those from the HerMES ``RARE'' source catalog, which allows for an easier identification of archival data.}
\tablenotetext{\rm b}{Redshift measured from 3\,mm line scans with the
  Combined Array for Research in Millimeter-wave Astronomy
  (CARMA). The CARMA spectra of HeLMS-5 and HXMM-01 only showed single
  CO lines. The redshifts for these source were confirmed from 2\,mm
  measurements with ALMA/ACA and NOEMA, respectively (D.\ Riechers et
  al., in preparation; see also Fu et al.\ \citeyear{fu13} for more
  details on HXMM-01). The redshift of HeLMS-5 reported here is
  updated from an incorrect value reported by Nayyeri et
  al.\ (\citeyear{nayyeri16}), which was based on a single CO
  line and a submillimeter photometric redshift. Orochi was also
  discovered independently from the HerMES survey by Ikarashi et
  al.\ (\citeyear{ikarashi11}).}
\tablenotetext{\rm c}{Redshift measured from 1\,cm line scans with the
  Zpectrometer on the Green Bank Telescope (GBT), showing a single CO
  line per source, and confirmed with CARMA (A. Harris et al., in
  preparation; D. Riechers et al., in preparation).}
\tablenotetext{\rm d}{\eco\ observations. Beam at 520\,$\mu$m.}
\tablenotetext{\rm e}{Beam at 310\,$\mu$m near the OH$^+$(1$_2$$\to$0$_1$) line.}
\tablenotetext{\rm f}{Redshift determined at 1\,cm and 3\,mm with the
  GBT, CARMA, and the Large Millimeter Telescope (LMT; Geach et
  al.\ \citeyear{geach15}; Harrington et al.\ \citeyear{harrington16};
  Su et al.\ \citeyear{su17}).}
\tablenotetext{\rm g}{Redshift measured from 3\,mm line scans with
  ALMA, and in the case of HeLMS-65, confirmed with the NSF’s Karl
  G.\ Jansky Very Large Array (VLA; Riechers et
  al.\ \citeyear{riechers17}; D. Riechers et al., in preparation; see
  also Asboth et al.\ \citeyear{asboth16}).}
\tablenotetext{\rm h}{Observed with the ALMA 12\,m array in the C43-5 configuration, covering 15\,m--1.4\,km baselines.}
\tablenotetext{\rm i}{Observed with PdBI/NOEMA in a special configuration between C and D.}
\end{deluxetable}
\vspace{-9mm}

\end{figure*}

\subsection{ALMA Observations}

We observed the \ico\ and OH$^+$(1$_1$$\to$0$_1$) lines ($\nu_{\rm
  rest}$=1036.9124 and 1033.0582\,GHz) toward 13 galaxies in our
sample at $z$=2.29--4.37 in bands 5--7 with the ALMA/ACA, using 10--12
7\,m antennas covering 8.9--48.9\,m baselines (see Tab.~\ref{t1}). For
four galaxies, the redshifts allowed us to also cover the
OH$^+$(1$_2$$\to$0$_1$) line ($\nu_{\rm rest}$=971.8053\,GHz) in the
same setups. Observations were carried out in cycles 4 and 6 under
good to excellent weather conditions on 2017 September 07 and between
2018 October 16 and November 27 (project IDs:\ 2016.2.00105.S and
2018.1.00922.S; PI:\ Riechers), spending between 15 and 50\,min on
source for different targets depending on their expected line
strengths. We also report \eco\ observations ($\nu_{\rm
  rest}$=576.2679\,GHz) for one target, carried out for 18\,min on
2018 December 04 as part of the second program. Nearby radio quasars
were observed for complex gain, bandpass, and absolute flux
calibration, except for the two 2017 observing runs, which used Uranus
as a flux calibrator (see Table \ref{t1}; the bandpass and flux
calibrators were identical where not listed separately). The absolute
flux calibration is estimated to be reliable to within $<$10\%.

The correlator was set up with two spectral windows of 1.875\,GHz
bandwidth (dual polarization) each per sideband, at a sideband
separation of 8\,GHz for all band 5 and 7 observations and typically
10\,GHz for all band 6 observations (with some limited modifications
to cover additional spectral lines in some cases). A spectral
resolution of 31.25\,MHz at a channel spacing of 15.625\,MHz was
chosen for all observations to reduce calibration overheads. Thus,
neighboring channels in spectra shown at full resolution are not
independent. In the current analysis, we only include contiguous
spectral windows that contain OH$^+$ lines, and we defer the remaining
part of the data to a future publication on other molecular species.

\begin{figure*}%[tbh!]
\epsscale{1.18}
\plotone{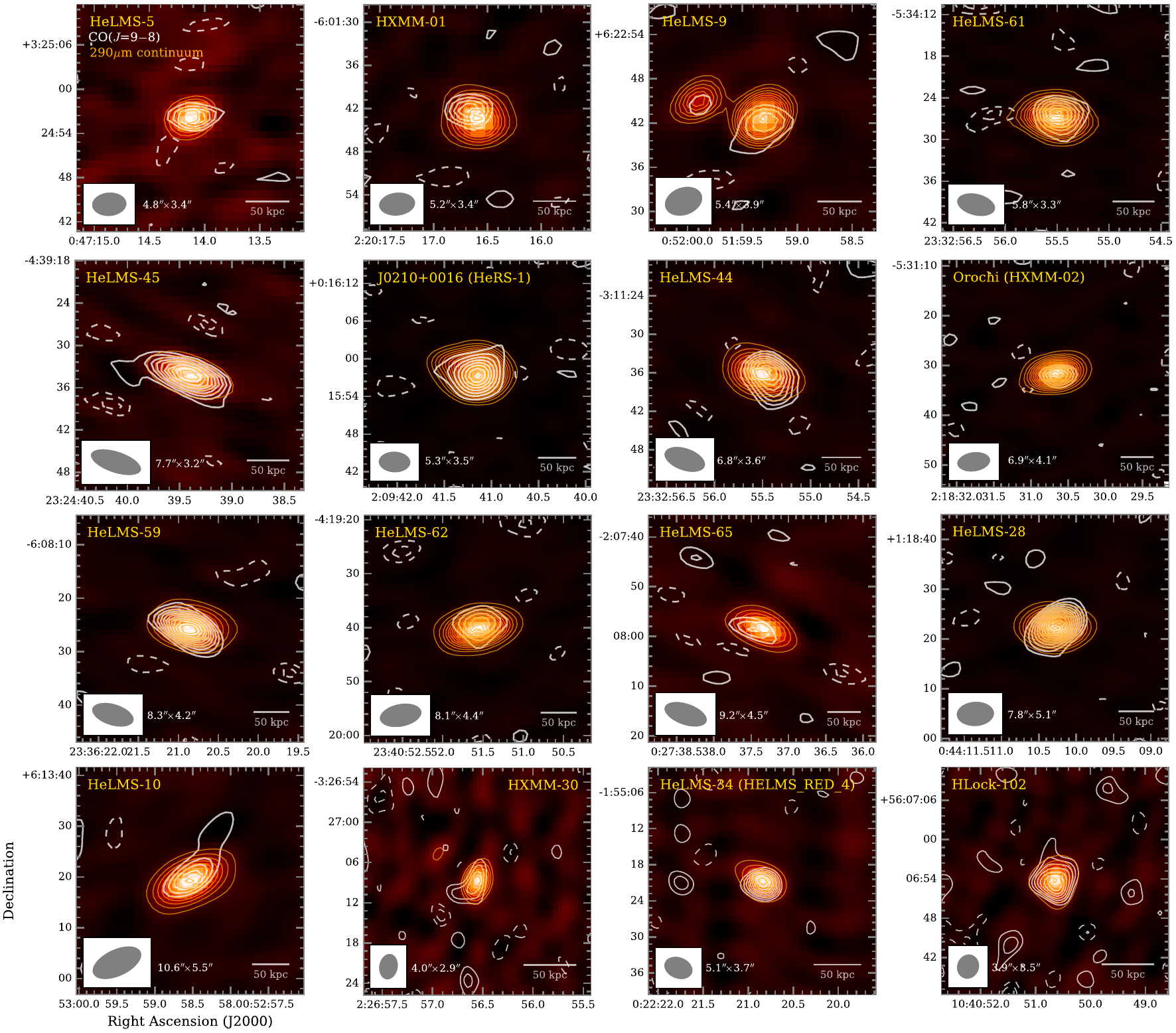}
\vspace{-5mm}

\caption{Rest-frame 290\,$\mu$m dust continuum maps and contours
  (orange), overlaid with \ico\ moment-0 contours (white) for the full
  sample, ordered with increasing redshift (last three panels are
  NOEMA observations, the remainder from ALMA/ACA). CO and dust
  emission is significantly detected in all sources, and co-spatial to
  within the positional uncertainties (except for HXMM-01, where the
  CO emission may be dominated by the northern merger component). Dust
  continuum contours start at $\pm$4$\sigma$, and are shown in steps
  of $\pm$4$\sigma$, where 1$\sigma$=1.28, 0.42, 0.52, 0.39, 0.80,
  0.90, 0.68, 0.33, 0.42, 0.45, 0.32, 0.26, 0.56, 0.27, 0.29, and
  0.26\,mJy\,beam$^{-1}$, respectively, except for except for HeLMS-5
  and 9 and XMM-30, where contour steps are $\pm$2$\sigma$. CO
  contours start at $\pm$2$\sigma$ and are shown in steps of
  $\pm$1$\sigma$=1.33, 1.08, 0.67, 0.39, 1.36, 0.84, 0.95, 0.56, 0.59,
  1.87, 0.42, 0.57, 0.80, 0.41, 0.52, and 0.67\,Jy\,\kms,
  respectively, except for J0210+0016, where contour steps are
  $\pm$2$\sigma$. The synthesized beam of the dust continuum map is
  shown in the bottom left corner of each panel, and the size is
  given. \label{f1}}
%\vspace{-5mm}
%
\end{figure*}

\begin{figure*}%[tbh]
\epsscale{1.18}
\plotone{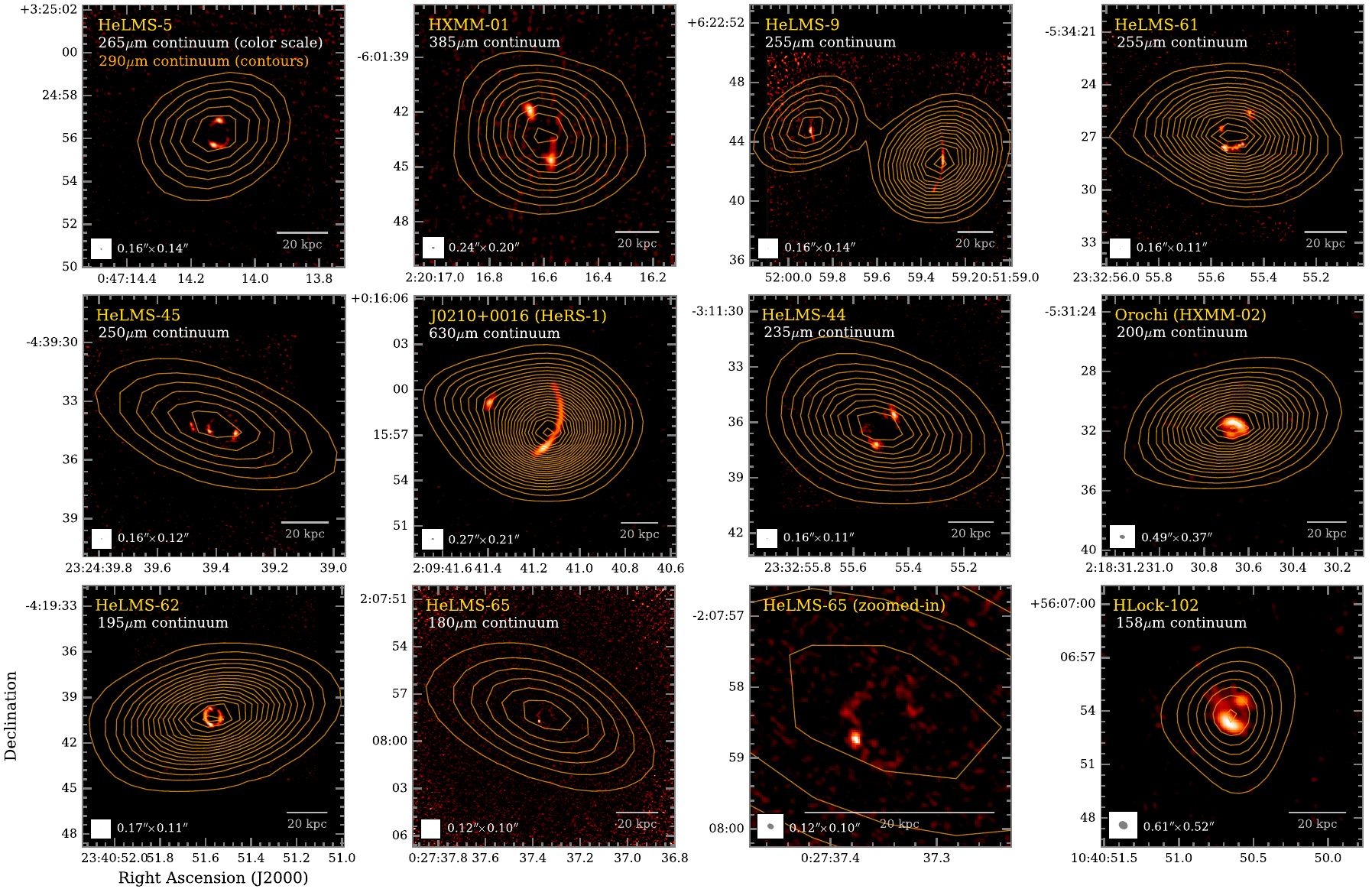}
\vspace{-5mm}

\caption{Same dust continuum contours as in Fig.~\ref{f1} (orange),
  but overlaid on ALMA (all except HLock-102) or SMA dust continuum
  imaging at higher spatial resolution for eleven of the sources in
  our sample (color scale; data adopted from Bussmann et
  al.\ \citeyear{bussmann15}; Oteo et al.\ \citeyear{oteo17};
  Amvrosiadis et al.\ \citeyear{amvrosiadis18}; Geach et
  al.\ \citeyear{geach18}; Xue et al.\ \citeyear{xue18}; Greenslade et
  al.\ \citeyear{greenslade20}; D.\ Riechers et al., in prep., but
  calibrated and imaged independently for most sources), based on
  availability in the literature. A second panel, zoomed-in by a
  factor of 4.5, is shown for HeLMS-65. The dust emission in the
  ALMA/ACA images is only substantially resolved for HeLMS-9 and
  J0210+0016, but strong lensing features are seen in all
  high-resolution images except for HXMM-01.  The synthesized beam of
  the high-resolution images is shown in the bottom left corner of
  each panel, and the size is given. Increased noise towards the edges
  of some fields is due to the primary beam correction.\label{f2}}
%\vspace{-5mm}
%
\end{figure*}

Data reduction was performed using version 5.4.0 or 5.6.1 of the {\sc
  casa} package (McMullin et al.\ \citeyear{mcmullin07}), aided by the
calibration pipeline included with each version. Data were mapped
manually using the CLEAN algorithm via the {\tt tclean} task with
``natural'' baseline weighting, resulting in the synthesized beam
sizes listed in Table~\ref{t1}. The main modification to the default
pipeline was to reduce the edge channel flagging in some cases where
they contained critical information, while still excluding channels
with poor calibration. The rms noise values for the line and continuum
emission in each data set are provided in the captions of
Figs.~\ref{f1} to \ref{f5}, and continuum-subtracted spectra are
provided in Figs.~\ref{f6} to \ref{f9}.\\[3mm]

%\clearpage
\begin{turnpage}

\begin{figure*}
\begin{deluxetable}{ l c c c c c c c c c c c c c }
  %\vspace{-7mm}

\tabletypesize{\scriptsize}
\tablecaption{Line and continuum parameters.\label{t2}}
\tablehead{
  Name & redshift & $I_{\rm CO(9-8)}$ & dv$_{\rm CO (9-8)}$ & $I_{\rm OH+(11-01)}^{\rm abs}$ & dv$_{\rm OH+(11-01)}^{\rm abs}$ & $I_{\rm OH+(11-01)}^{\rm em}$ & dv$_{\rm OH+(11-01)}^{\rm em}$ & $I_{\rm OH+(12-01)}^{\rm abs}$ & dv$_{\rm OH+(12-01)}^{\rm abs}$ & $I_{\rm OH+(12-01)}^{\rm em}$ & dv$_{\rm OH+(12-01)}^{\rm em}$ & $S_{290}$ & $S_{310}$ \\
 & & (Jy \kms ) & (\kms ) &  (Jy \kms ) & (\kms ) &  (Jy \kms ) & (\kms ) &  (Jy \kms ) & (\kms ) &  (Jy \kms ) & (\kms ) & (mJy) & (mJy) }
\startdata
HeLMS-5    & 2.2919 & 9.1$\pm$1.4 & 596$\pm$81 & --3.36$\pm$0.65 & 186$\pm$34 & {\em 1.55$\pm$0.62} & {\em 386$\pm$145} & & & & & 25.4$\pm$1.7 & \\
HXMM-01    & 2.3079 & 7.3$\pm$1.2 & 600$\pm$55 & {\em --1.38$\pm$0.47} & {\em 265$\pm$72} & & & & & & & 27.92$\pm$0.97 & \\
HeLMS-9\tablenotemark{a} & 2.3934 & 6.7$\pm$1.3 & 694$\pm$90 & {\em --0.81$\pm$0.29} & {\em 292$\pm$78} & 3.7$\pm$1.1(e) & 1248$\pm$235 & & & & & 29.6$\pm$1.7 & \\
           &        &             & &                 & & 1.16$\pm$0.33(w) & 531$\pm$90 & & & & &              & \\
HeLMS-61   & 2.4022 & 3.81$\pm$0.48 & 100$\pm$8 & & & 2.89$\pm$0.40 & 199$\pm$23 & & & & & 32.89$\pm$0.78 & \\
           &        &               & 52$\pm$14 &                 & & & & & & & &              & \\
HeLMS-45   & 2.4726 & 18.9$\pm$2.7 & 610$\pm$31 & --2.90$\pm$0.55 & 191$\pm$40 & {\em 2.01$\pm$0.49(ne)} & {\em 160$\pm$35} & & & & & 22.3$\pm$1.2 & \\
           &        &              & &                 & & 1.98$\pm$0.62(sw) & 251$\pm$62 & & & & &              & \\
J0210+0016 & 2.5534 & 19.2$\pm$1.7 & 146$\pm$12 & & & 11.6$\pm$1.8 & 272$\pm$38 & & & & & 149.4$\pm$6.2 & \\
           &        &              & 490$\pm$56 & & &              & 797$\pm$210 & & & & &              & \\
HeLMS-44   & 2.6895 & 11.5$\pm$1.9 & 648$\pm$72 & --6.93$\pm$0.88 & 573$\pm$59 & 2.56$\pm$0.76 & 591$\pm$194 & & & & & 39.3$\pm$1.1 & \\
Orochi     & 3.3903 & 4.03$\pm$0.61 & 505$\pm$103 & --2.84$\pm$0.47 & 259$\pm$31 & & & --2.76$\pm$0.39 & 215$\pm$30 & & & 24.09$\pm$0.54 & 17.72$\pm$0.35 \\
HeLMS-59   & 3.4346 & 8.9$\pm$1.0 & 708$\pm$44 & {\em (not covered)} & & $>$1.66$\pm$0.33\tablenotemark{c} & 494$\pm$205\tablenotemark{c} & --1.40$\pm$0.48 & 247$\pm$29 & 2.30$\pm$0.62 & 613$\pm$97\tablenotemark{e} & 25.41$\pm$0.95 & 21.2$\pm$1.5 \\
HeLMS-62   & 3.5027 & 11.8$\pm$2.3 & 380$\pm$47 & --5.59$\pm$0.50 & 683$\pm$81 & {\em (not covered)} & & --4.77$\pm$0.78 & 337$\pm$28 & 5.2$\pm$1.2 & 933$\pm$120 & 32.15$\pm$0.43 & 21.79$\pm$0.90 \\
           &        &              & 1114$\pm$310 &                 & & & & & & & &              & \\
HeLMS-65   & 3.7966 & 1.61$\pm$0.38 & 201$\pm$37 & {\em --0.34$\pm$0.14} & {\em 89$\pm$30} & 1.66$\pm$0.51 & 363$\pm$92 & & & & & 8.66$\pm$0.28 & \\
           &        &               & 222$\pm$83 &                 & & & & & & & &              & \\
HeLMS-28   & 4.1625 & 8.36$\pm$0.70 & 776$\pm$59 & --3.33$\pm$0.37 & 324$\pm$38 & & & --1.69$\pm$0.34 & 524$\pm$125\tablenotemark{d} & & & 18.40$\pm$0.24 & 14.66$\pm$0.37 \\
HeLMS-10   & 4.3726 & 4.72$\pm$0.84 & 360$\pm$52 & --1.36$\pm$0.42 & 85$\pm$104 & & & & & & & 21.82$\pm$0.99 & \\
HXMM-30    & 5.094  & 2.18$\pm$0.44 & 307$\pm$101 & --3.45$\pm$0.61 & 999$\pm$347 & & & & & & & 5.81$\pm$0.45 & \\
HeLMS-34   & 5.1614 & 5.82$\pm$0.56 & 632$\pm$90 & --2.68$\pm$0.58 & 643$\pm$321 & & & & & & & 7.88$\pm$0.55 & \\
HLock-102  & 5.2915 & 9.7$\pm$1.5 & 777$\pm$99 & --1.35$\pm$0.27 & 132$\pm$40 & 3.45$\pm$0.64 & 645$\pm$280 & & & & & 10.27$\pm$0.66 & \\
ADFS-27\tablenotemark{b} & 5.6550 & 1.80$\pm$0.04 & 597$\pm$18 & --2.96$\pm$0.027 & 422$\pm$50 & 0.147$\pm$0.018 & 745$\pm$87 & {\em (not covered)} & & 0.17$\pm$0.04 & $\sim$740 & 2.67$\pm$0.05 & 2.09$\pm$0.05 \\
HFLS3\tablenotemark{b} & 6.3369 & 2.77$\pm$0.45 & 497$\pm$107 & --0.56$\pm$0.18 & 416$\pm$170 & $>$0.90\tablenotemark{c} & $>$765 & & & & & 3.22$\pm$0.12 & 2.38$\pm$0.11 \\
\enddata
\tablecomments{ Tentative detections are indicated in italic font.}
\tablenotetext{\rm a}{The \eco\ FWHM line width is dv$_{\rm CO(5-4)}$=700$\pm$64\,\kms. This shows that the broadened component is due to OH$^+$ emission, not \ico. The 520\,$\mu$m continuum flux is $S_{520}$=3.69$\pm$0.48\,mJy.}
\tablenotetext{\rm b}{Adopted from Riechers et al.\ (\citeyear{riechers13b,riechers20c}).}
\tablenotetext{\rm c}{Component not fully covered by the bandpass.}
\tablenotetext{\rm d}{Fit to the line profile includes a moderate signal-to-noise ratio component, and thus, is less reliable than that to the OH$^+$(1$_1$$\to$0$_1$) line.}
\tablenotetext{\rm e}{Line may include a broad wing, which if confirmed would increase the best estimate to 995$\pm$171\,\kms. If real, this would likely also imply some contribution of OH$^+$(1$_1$$\to$0$_1$) emission to the measured \ico\ line flux.}
\end{deluxetable}
\vspace{-9mm}

\end{figure*}

%\clearpage
\end{turnpage}

\subsection{NOEMA Observations}

We observed the \ico\ and OH$^+$(1$_1$$\to$0$_1$) lines toward 3
galaxies in our sample at $z$=5.09--5.29 in band 2 with the NOrthern
Extended Millimeter Array (NOEMA). Observations were carried out with
5 or 6 antennas in the most compact D configuration under good weather
conditions between 2013 June 12 and August 17 and between 2015
September 7 and 21 (project IDs:\ X045 and S15CY; PI:\ Riechers),
spending between 110 and 470\,min on source. Nearby radio quasars were
observed for complex gain, bandpass, and absolute flux calibration
(see Table \ref{t1}), resulting in a flux scale that is reliable
within $<$10\%. All observations were carried out with the WideX
correlator, providing 3.6\,GHz bandwidth (dual polarization) at 2\,MHz
spectral resolution.

Data reduction was performed using the {\sc gildas} package. Data were
mapped using the CLEAN algorithm with ``natural'' baseline weighting,
resulting in the synthesized beam sizes listed in Table~\ref{t1}. The
rms noise values for the line and continuum emission in each data set
are provided in the captions of Figs.~\ref{f1} and \ref{f3}, and
continuum-subtracted spectra are provided in Figs.~\ref{f6} and
\ref{f7}.

\section{Results}

\subsection{Continuum Emission}

We successfully detect rest-frame 290\,$\mu$m continuum emission at
high peak signal-to-noise ratios (SNRs) of 16--98 toward all 16
galaxies in the sample (Fig.~\ref{f1}). We also detect rest-frame
310\,$\mu$m emission in all four galaxies studied at this wavelength
(Fig.~ \ref{f4}). We further detect rest-frame 520\,$\mu$m emission
toward HeLMS-9 as part of the \eco\ observations reported here
(Fig.~\ref{f5}). The dust continuum emission is only marginally
resolved at most in all sources except the large Einstein radius
lenses HeLMS-9 and J0210+0016, and only the former is fully resolved
into two components corresponding to its lensing arcs
(Fig.~\ref{f2}). Continuum fluxes were extracted from two-dimensional
Gaussian fitting in the image plane for all sources observed with
ALMA, and from two-dimensional Gaussian fitting in the visibility
plane for the sources observed with NOEMA (given the near-equatorial
declination and resulting limited uv coverage for two of the
sources). Two Gaussian components were used for HeLMS-9, and the flux
of both components was added by propagating the errors of each
component. All continuum fluxes and uncertainties from the fitting are
reported in Table~\ref{t2}. Given the high SNRs of all detections,
uncertainties are typically dominated by the absolute flux
calibration, which is accounted for in our subsequent
analysis. Comparison fluxes at standard wavelengths are provided in
Table~\ref{t3}.

\begin{figure*}%[tbh]
\epsscale{1.18}
\plotone{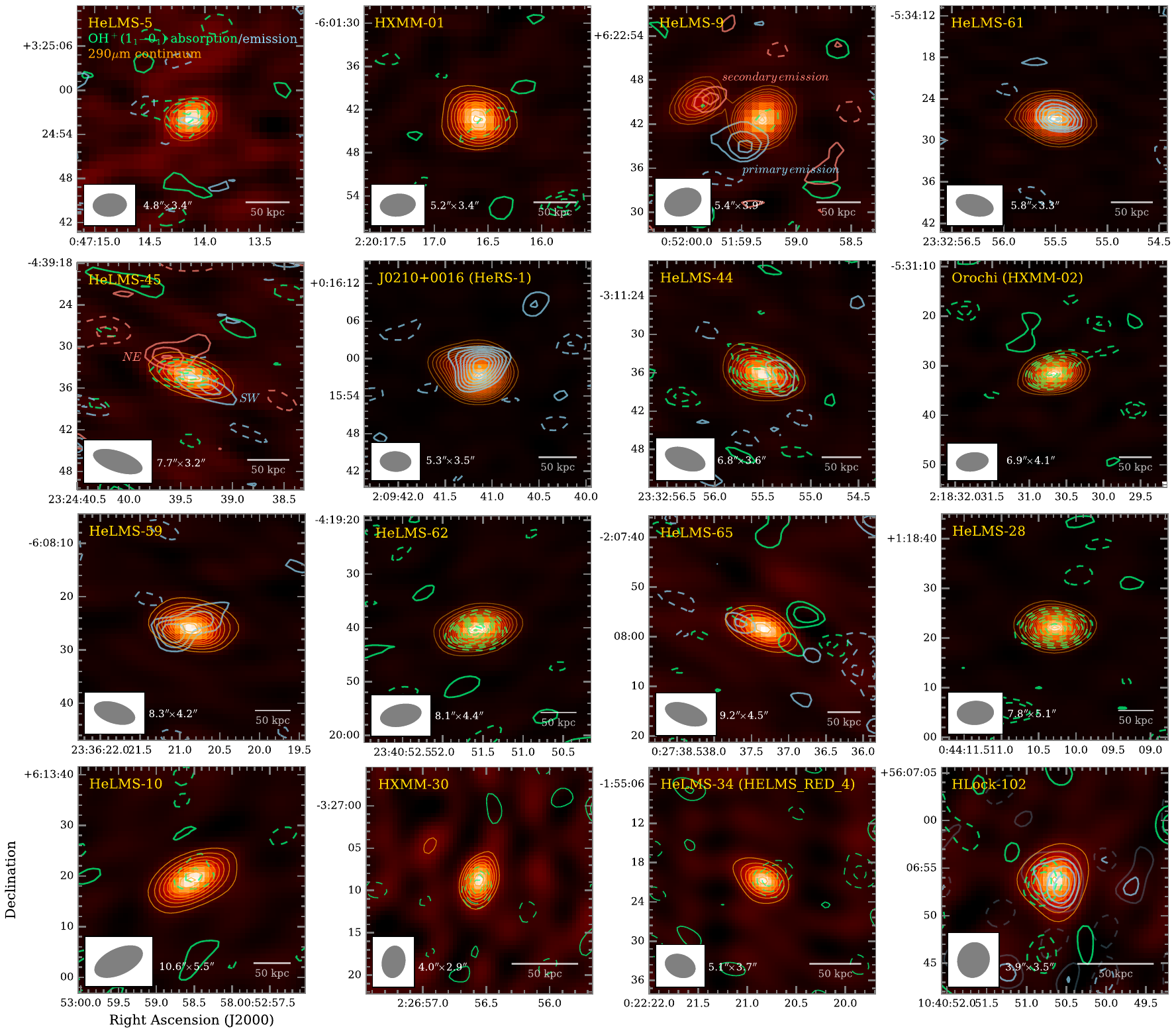}
\vspace{-5mm}

\caption{Same continuum emission as in Fig.~\ref{f1}, but showing
  OH$^+$(1$_1$$\to$0$_1$) instead of \ico. OH$^+$ absorption is
  displayed in dashed green contours, and OH$^+$ emission is displayed
  in blue contours where one component is detected, and in blue/red
  contours for bluer/redder components where multiple components are
  detected. OH$^+$ absorption contours are shown for all sources
  except HeLMS-61, HeRS-1, and HeLMS-59, and start at $\pm$2$\sigma$,
  progressing in steps of $\pm$1$\sigma$=0.71, 0.47, 0.26, 0.49, 0.63,
  0.27, 0.49, 0.15, 0.27, 0.38, 0.57, 0.47, and 0.26\,Jy\,\kms,
  respectively (the expected absorption component for HeLMS-59 fell
  outside the tuning range). OH$^+$ emission contours are shown for
  HeLMS-5, 9, 61, 45, J0210+0016, HeLMS-44, 59, 65, and HLock-102, and
  start at $\pm$2$\sigma$, progressing in steps of
  $\pm$1$\sigma$=0.58, 0.30/0.63 (blue/red primary/secondary
  components), 0.37, 0.46/0.47 (blue/red south-west/north-east
  components), 0.82, 0.64, 0.35, 0.09, and 0.38\,Jy\,\kms,
  respectively (the expected emission component for HeLMS-62 fell
  outside the tuning range). OH$^+$ absorption is detected or
  tentatively detected in 14/16 sources (see Fig.~\ref{f4} for
  HeLMS-59), and emission is detected or tentatively detected in 10/16
  sources (see Fig.~\ref{f4} for HeLMS-62). Only 2/16 sources show
  OH$^+$ emission but no absorption.\label{f3}}
%\vspace{-5mm}
%
\end{figure*}

\begin{figure*}%[tbh]
\epsscale{1.18}
\plotone{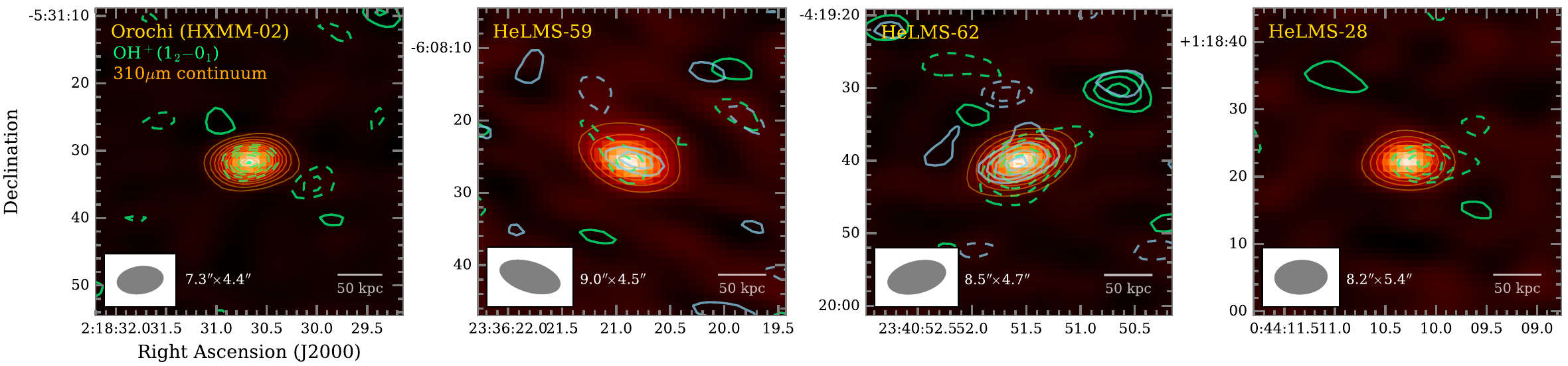}
\vspace{-5mm}

\caption{Same as Fig.~\ref{f3}, but showing 310\,$\mu$m dust continuum
  (color scale and orange contours) and OH$^+$(1$_2$$\to$0$_1$) for
  the sources where the latter line was covered. Dust continuum
  contours are shown in steps of $\pm$4$\sigma$, where 1$\sigma$=0.34,
  0.78, 0.58, and 0.52\,mJy\,beam$^{-1}$, respectively (left to
  right). OH$^+$ absorption contours (green) start at $\pm$2$\sigma$
  and are shown in steps of $\pm$1$\sigma$=0.36, 0.38, 0.48, and
  0.31\,Jy\,\kms, respectively. OH$^+$ emission contours (blue) for
  HeLMS-59 and 62 start at $\pm$2$\sigma$ and are shown in steps of
  $\pm$1$\sigma$=0.58 and 0.91\,Jy\,\kms, respectively.\label{f4}}
%\vspace{-5mm}
%
\end{figure*}

\begin{figure*}%[tbh]
\epsscale{1.18}
\plotone{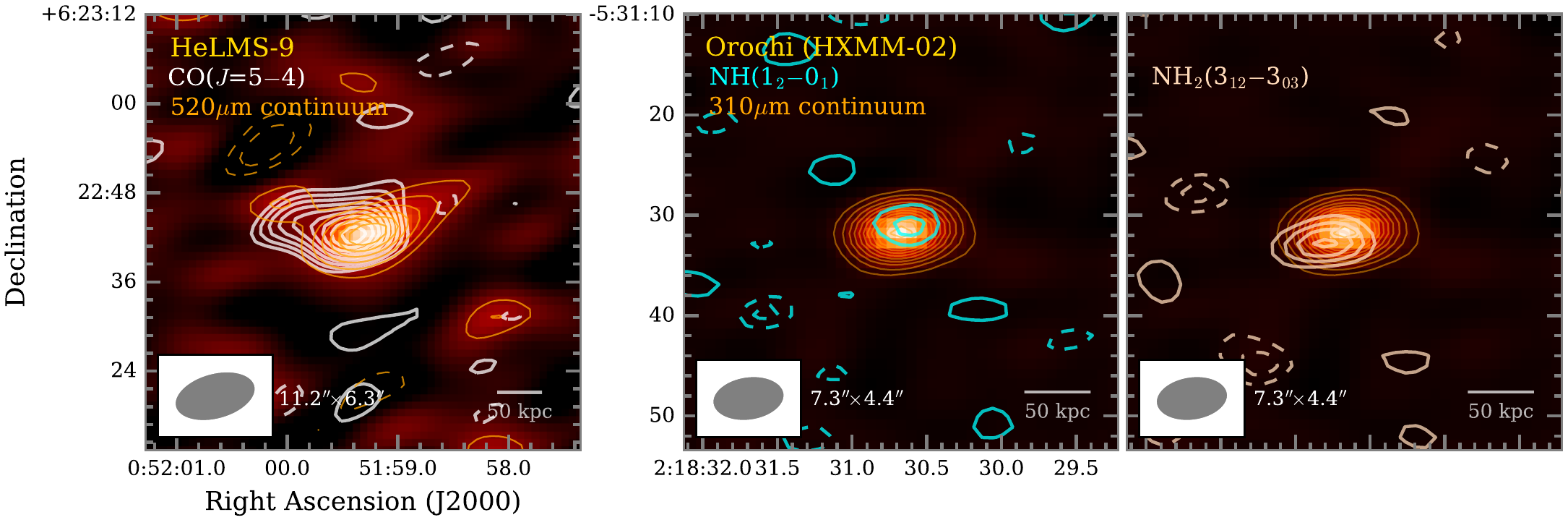}
\vspace{-5mm}

\caption{Rest-frame 520\,$\mu$m dust continuum map and contours (orange), overlaid with \eco\ moment-0 contours (white) toward HeLMS-9 (left), and same dust continuum contours as in Fig.~\ref{f4} for Orochi, overlaid with NH (middle; light blue) and NH$_2$ (right; light red) contours. Continuum contours for HeLMS-9 are shown in steps of $\pm$1$\sigma$=0.32\,mJy\,beam$^{-1}$, starting at $\pm$2$\sigma$. Line contours are shown in steps of $\pm$1$\sigma$=0.93, 0.40, and 0.44\,Jy\,\kms, respectively (left to right), starting at $\pm$2$\sigma$. The synthesized beam of the dust continuum map is shown in the bottom left corner of each panel, and the size is given. \label{f5}}
%\vspace{-5mm}
%
\end{figure*}

\begin{figure*}%[tbh]
\epsscale{1.18}
\plotone{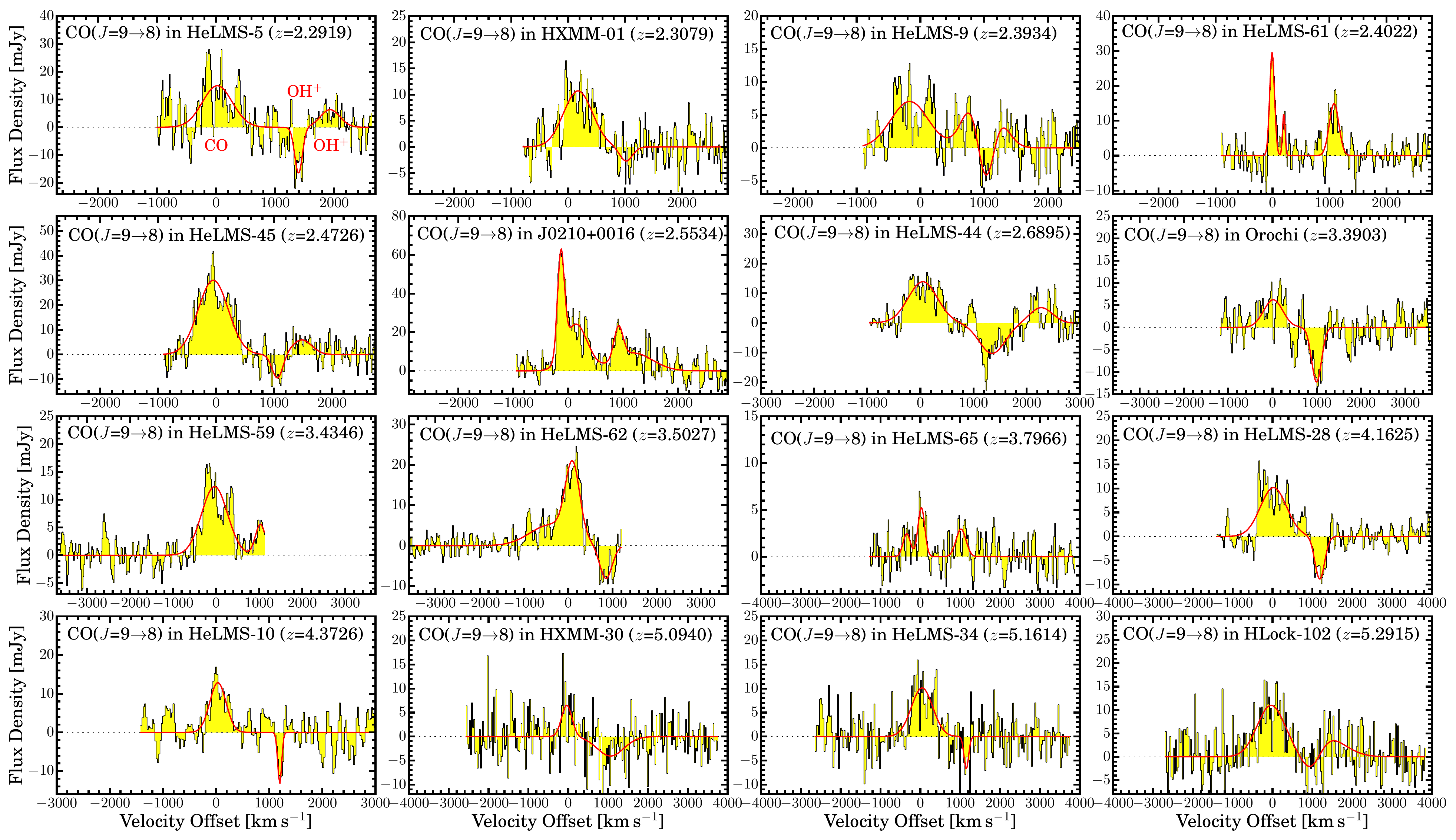}
\vspace{-5mm}

\caption{Spectra of \ico\ emission lines (histograms) and Gaussian fits (red curves) to the line spectra, shown at 15.625\,MHz spectral resolution. All spectra were extracted centered on the emission peaks in the moment 0 maps found after continuum subtraction (Fig.~\ref{f1}). Significantly redward features are due to OH$^+$ absorption/emission (see Fig.~\ref{f7}). All features were fitted simultaneously, but only the fit parameters for \ico\ were adopted.\label{f6}}
%\vspace{-5mm}
%
\end{figure*}

\begin{figure*}%[tbh]
\epsscale{1.18}
\plotone{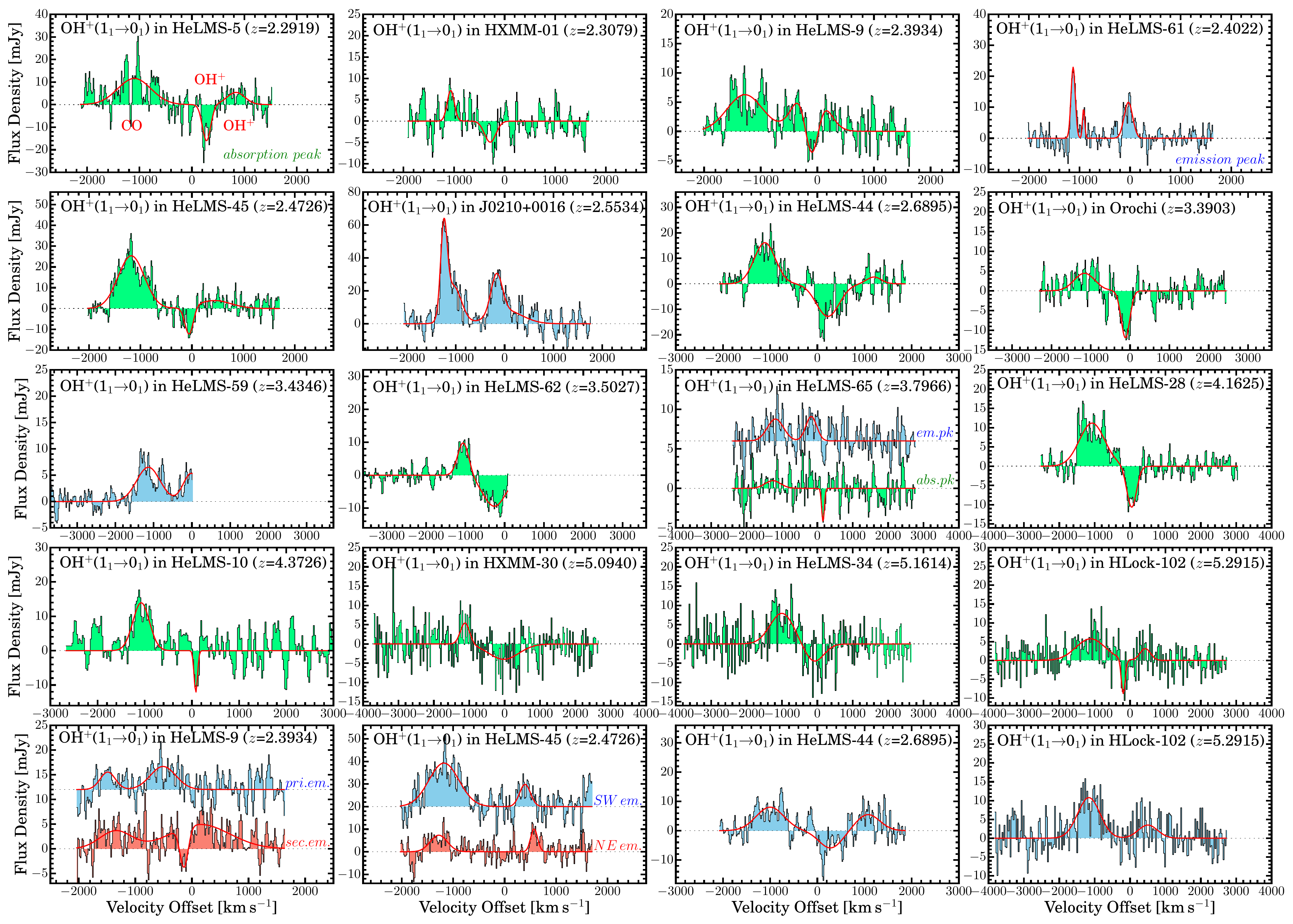}
\vspace{-5mm}

\caption{Spectra of OH$^+$(1$_1$$\to$0$_1$) absorption/emission lines (histograms) and Gaussian fits (red curves) to the line spectra, shown at 15.625\,MHz spectral resolution. All spectra were extracted centered on the absorption (green) or emission peaks (blue) in the moment 0 maps found after continuum subtraction (Fig.~\ref{f3}). Red histograms are used for the redder emission components in sources that show multiple emission components (i.e., the component toward the fainter lensed image of HeLMS-9, and the north-eastern (NE) emission component toward HeLMS-45). Secondary components are offset on the $y$ axis for clarity. The same color coding is used as in the contour maps in Fig.~\ref{f3}. All features were fitted simultaneously, but only the fit parameters for OH$^+$ were adopted.\label{f7}}
\vspace{-5mm}

\end{figure*}

\begin{figure*}%[tbh]
\epsscale{1.18}
\plotone{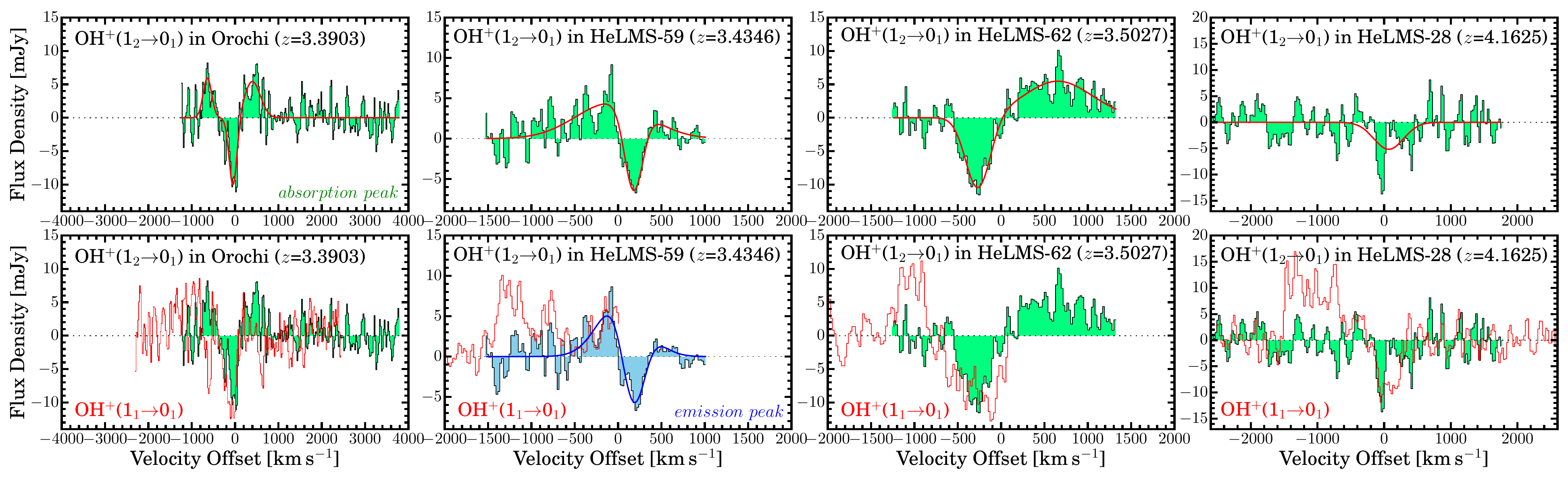}
\vspace{-5mm}

\caption{Spectra of OH$^+$(1$_2$$\to$0$_1$) absorption/emission lines (histograms) and Gaussian fits (red curves) to the line spectra (top) and comparison to the OH$^+$(1$_1$$\to$0$_1$) spectra (bottom), all shown at 15.625\,MHz spectral resolution. All spectra were extracted centered on the absorption peaks in the moment 0 maps found after continuum subtraction (Fig.~\ref{f4}), except the bottom panel for HeLMS-59, which was extracted centered on the emission peak (where the blue curve shows a Gaussian fit to this spectrum). The same color coding is used as in the contour maps in Fig.~\ref{f4}. \ico\ emission was not subtracted from the comparison spectra. Bottom left:\ The OH$^+$ 1$_2$$\to$0$_1$  and 1$_1$$\to$0$_1$ absorption features in Orochi show comparable line profiles, but the former spectrum shows additional emission at velocities where none is seen in the latter. The emission thus is unlikely to be due to OH$^+$ (see Fig.~\ref{f9} for line identifications). All features were fitted simultaneously, but only the fit parameters for OH$^+$ were adopted.\label{f8}}
%\vspace{-5mm}
%
\end{figure*}

\begin{figure*}%[tbh!]
\epsscale{1.18}
\plotone{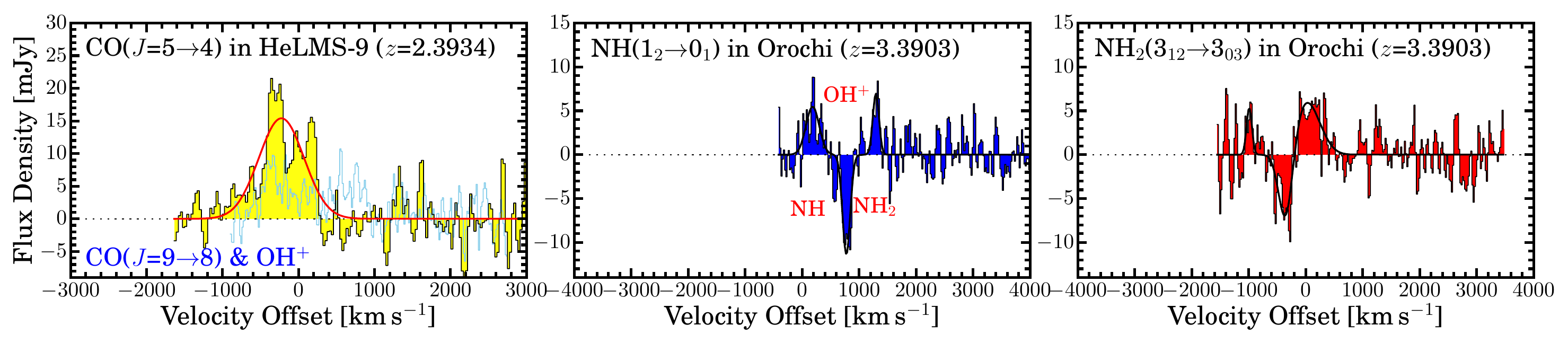}
\vspace{-5mm}

\caption{Spectra of emission lines other than \ico\ or OH$^+$, shown at 15.625\,MHz spectral resolution, and Gaussian fits (red curves) to the line spectra. All spectra were extracted centered on the emission peaks in the moment 0 maps found after continuum subtraction (Fig.~\ref{f5}). Left:\ The overlay of the \eco\ (yellow histogram) and \ico\ (light blue) line spectra toward HeLMS-9 shows that the redward emission component observed in the latter is not seen in the former, and thus, most likely due to blueshifted OH$^+$ emission. This is consistent with the spectral decompositions shown in Figs.~\ref{f6} and \ref{f7}, which assume that the emission at +400--1000\,\kms\ is associated with OH$^+$, not CO. Middle and right:\ The emission components in the OH$^+$(1$_2$$\to$0$_1$) spectrum are consistent with redshifted NH emission (middle) and NH$_2$ emission at the systemic velocity (right). All features were fitted simultaneously, but only the fit parameters for NH (middle) or NH$_2$ (right) were adopted. \label{f9}}
%\vspace{-5mm}
%
\end{figure*}

\begin{figure*}
\begin{deluxetable}{ l c c c c c c c c }
  %\vspace{-7mm}

\tabletypesize{\scriptsize}
\tablecaption{Line shifts, and OH$^+$ optical depths, column densities, and emission luminosities.\label{t2b}}
\tablehead{
  Name & redshift & v$_0^{\rm OH+,abs}$--v$_0^{\rm CO}$ & v$_0^{\rm OH+,em}$--v$_0^{\rm CO}$ & origin & $\tau_{\rm OH+}$ & $\tau_{\rm OH+}$dv & $N$(OH$^+$)\tablenotemark{a} & $L'^{\rm \,em}_{\rm OH^+(1-0)}$\tablenotemark{b} \\
 & & (\kms ) & (\kms ) & & & (\kms ) & (10$^{14}$\,cm$^{-2}$) & (10$^{10}$\,$L_l$ ) }
\startdata
HeLMS-5    & 2.2919 & 258$\pm$33    & {\em (795$\pm$63)} & ambiguous & 1.3$^{+0.7}_{-0.4}$ & 343$^{+186}_{-109}$ & 16.7$^{+9.0}_{-5.3}$ & {\em 0.50$\pm$0.20} \\
HXMM-01    & 2.3079 & {\em (--457$\pm$36)} & & ambiguous & {\em 0.32$^{+0.14}_{-0.12}$} & {\em 116$^{+50}_{-44}$} & {\em 5.6$^{+2.4}_{-2.1}$} & \\
HeLMS-9    & 2.3934 & {\em (68$\pm$35)} & --341$\pm$48(e) & ambiguous & {\em 0.25$^{+0.10}_{-0.11}$} & {\em 53$^{+23}_{-21}$} & {\em 2.6$^{+1.1}_{-1.0}$} & 1.69$\pm$0.40 \\
           &        &               & 313$\pm$77(w) & & & & & \\
HeLMS-61   & 2.4022 &               & --26$\pm$10 & inflow? & & & & 1.01$\pm$0.14 \\
           &        &               & --232$\pm$11 & & & & & \\
HeLMS-45   & 2.4726 & (--6$\pm$17)    & {\em (622$\pm$18)(ne)} & outflow & 0.55$^{+0.15}_{-0.13}$ & 173$^{+47}_{-41}$ & 8.4$^{+2.3}_{-2.0}$ & 1.47$\pm$0.29 \\
           &        &               & 397$\pm$25(sw) & & & & & \\
J0210+0016 & 2.5534 &               & --32$\pm$15 & inflow? & & & & 4.52$\pm$0.70 \\
HeLMS-44   & 2.6895 & 177$\pm$36    & 1026$\pm$61 & outflow? & 0.30$^{+0.05}_{-0.04}$ & 179$^{+27}_{-26}$ & 8.7$^{+1.3}_{-1.3}$ & 1.09$\pm$0.32 \\
Orochi     & 3.3903 & --132$\pm$41  & & outflow & 0.43$\pm$0.09 & 146$^{+32}_{-29}$ & 7.1$^{+1.5}_{-1.4}$ & \\
HeLMS-59\tablenotemark{c} & 3.4346 & 213$\pm$20   & 94$\pm$60 & inflow & 0.26$^{+0.11}_{-0.10}$ & 88$^{+36}_{-33}$ & 2.5$^{+1.0}_{-0.9}$ & 1.66$\pm$0.45 \\
HeLMS-62\tablenotemark{c} & 3.5027 & --363$\pm$15  & 561$\pm$37 & outflow & 0.46$^{+0.10}_{-0.09}$ & 199$^{+43}_{-39}$ & 11.1$^{+1.4}_{-1.3}$ & 3.88$\pm$0.89 \\
HeLMS-65   & 3.7966 & {\em (136$\pm$20)}    & --197$\pm$40 & inflow & {\em 1.88$^{+\infty}_{-1.19}$} & {\em 244$^{+\infty}_{-154}$} & {\em 11.9$^{+\infty}_{-7.5}$} & 1.25$\pm$0.38 \\
           &        & {\em (502$\pm$36)}    & & & & & & \\
HeLMS-28   & 4.1625 & (20$\pm$28)     & & ambiguous & 0.40$^{+0.06}_{-0.05}$ & 161$^{+23}_{-21}$ & 7.8$^{+1.1}_{-1.0}$ & \\
HeLMS-10   & 4.3726 & (49$\pm$49)     & & ambiguous & 0.57$^{+0.27}_{-0.21}$ & 83$^{+39}_{-31}$ & 4.0$^{+1.9}_{-1.5}$ & \\
HXMM-30    & 5.094  & (16$\pm$139)    & & ambiguous & 1.43$^{+0.83}_{-0.45}$ & 1109$^{+642}_{-346}$ & 54$^{+31}_{-17}$ & \\
HeLMS-34   & 5.1614 & (--133$\pm$150) & & ambiguous & 0.75$^{+0.27}_{-0.21}$ & 417$^{+153}_{-120}$ & 20.3$^{+7.5}_{-5.8}$ & \\
HLock-102  & 5.2915 & (--152$\pm$159) & 544$\pm$193 & outflow & 1.90$^{+\infty}_{-0.76}$ & 381$^{+\infty}_{-152}$ & 18.5$^{+\infty}_{-7.4}$ & 4.33$\pm$0.80 \\
ADFS-27\tablenotemark{d} & 5.6550 & --216$\pm$16 & 233$\pm$10 & outflow & 0.35$\pm$0.04 & 234$\pm$26 & 11.4$\pm$1.2 & 0.20$\pm$0.02 \\
HFLS3\tablenotemark{d} & 6.3369 & 179$\pm$74 & ---\tablenotemark{e} & ambiguous & 0.49$^{+0.21}_{-0.17}$ & 205$^{+87}_{-72}$ & 10.0$^{+4.2}_{-3.5}$ & $>$1.48 \\
\enddata
\tablecomments{Tentative detections are indicated in italic font. Values in parantheses are consistent with no velocity shifts (tentative detections are included in this category by default). ``Origin'' is the most likely interpretation based on the presence of P-Cygni (outflow) or inverse P-Cygni (inflow) profiles, or when not available or ambiguous, the blue- or redshift of the absorption and emission components.}
\tablenotetext{\rm a}{These values can be translated to approximate neutral HI column densities by assuming a relative abundance of [OH$^+$/H]=--7.8, i.e., by multiplying the listed values by a factor of 6.3$\times$10$^7$ (e.g., Bialy et al.\ \citeyear{bialy19}).}
\tablenotetext{\rm b}{Given in units of $L_l$=K\,\kms\,pc$^2$. Apparent values not corrected for gravitational magnification where applicable.}
\tablenotetext{\rm c}{Velocity shifts and $L'^{\rm \,em}_{\rm OH^+(1-0)}$ are reported relative to OH$^+$ 1$_2$$\to$0$_1$ instead of 1$_1$$\to$0$_1$ due to the incomplete coverage of the latter by our observations.}
\tablenotetext{\rm d}{Adopted from Riechers et al.\ (\citeyear{riechers13b,riechers20c}).}
\tablenotetext{\rm e}{Component not fully covered by the bandpass.}
\end{deluxetable}
\vspace{-9mm}

\end{figure*}

\subsection{Line Emission and Absorption}

\subsubsection{CO Emission}

We successfully detect \ico\ emission toward all 16 galaxies in the
sample at peak SNRs of 5--16 in the continuum-subtracted moment 0 maps
(except for HeLMS-65, which is detected at a peak SNR of 4.2;
Fig.~\ref{f1}). The continuum subtraction was carried out in the
visibility plane, and accounts for the spectral slope of the continuum
where measurable within the line-free spectral ranges. The line
emission is only marginally resolved at best in all cases except
HeLMS-9, which is resolved into two components. Line spectra were
extracted centered on the peak positions of the line-averaged emission
(Fig.~\ref{f6}).

Line parameters were obtained from Gaussian fitting to the CO line
profiles, using two Gaussian components where appropriate. Other lines
in the bandpass were fitted simultaneously with additional Gaussian
functions to include reliable estimates of the uncertainties of all
fitting parameters, and the integrated line fluxes were compared to
those found from two-dimensional Gaussian fitting to the emission in
the moment 0 maps to warrant internal consistency. The resulting line
fluxes are reported in Table~\ref{t2}, and line luminosities are given
in Table~\ref{t3}.

\subsubsection{OH$^+$ Emission and Absorption}

We successfully detect OH$^+$(1$_1$$\to$0$_1$) emission or absorption
toward all 16 galaxies in the sample, and OH$^+$(1$_2$$\to$0$_1$)
toward all 4 galaxies in which the line was covered (Figs.~\ref{f3}
and \ref{f4}). OH$^+$ absorption (emission) is detected or tentatively
detected in 14/16 (10/16) sources. 2/16 sources show OH$^+$ emission,
but no absorption. 6/16 sources show OH$^+$ absorption, but no
emission. 8/16 sources show evidence for both absorption and emission
components when including tentative detections. Among the absorption
lines, three are tentatively detected at peak SNRs of 2.4--2.9, two
are detected at moderate peak SNRs of 3.3, and the remaining 12 lines
are detected solidly at peak SNRs of 4.6--11 (all SNRs were estimated
based on the continuum-subtracted moment 0 maps). The peak postions of
the absorption features are always coincident with the continuum peak
positions within the relative uncertainties. Among the emission lines,
one is tentatively detected at a peak SNR of 2.5, five are detected at
moderate peak SNRs of 3.2--3.7, and seven are detected solidly at peak
SNRs of 4.1--9.1 (two sources, HeLMS-9 and 45, show multiple emission
components). All detections are plausible, but independent
confirmation is required for all features with SNR$<$4 (SNR$<$5 for
sources that show potential spatial offsets from the CO and continuum
peaks).

Line spectra were extracted centered on the peak positions of the
line-averaged emission (Fig.~\ref{f7} and \ref{f8}). Separate spectra
were extracted for absorption and emission peaks where the positions
may not coincide (HeLMS-44, 59, and 65, and HLock-102), and/or where
multiple emission components are detected (HeLMS-9 and 45 only). Most
of the offsets are marginal due to the moderate SNR of the emission
components (see discussion in Section 3.3) and the potential interplay
between emission/absorption components (such that OH$^+$ emission may
be partially masked due to foreground OH$^+$ absorption in some
velocity ranges; see more detailed discussion below). As such, while
our data may suggest that spatial offsets between absorption/emission
components such as previously observed in ADFS-27 (Riechers et
al.\ \citeyear{riechers20c}) may not be uncommon, we defer a more
general interpretation of absorption/emission peak position offsets to
future work on higher resolution data. Within the uncertainties, the
OH$^+$(1$_1$$\to$0$_1$) and OH$^+$(1$_2$$\to$0$_1$) lines show the
same profiles and velocities for all sources where both were covered,
with the possible exception of HeLMS-62, where the absorption
component of the former may be broader. The SNR of the
OH$^+$(1$_2$$\to$0$_1$) line in HeLMS-28 is too limited to warrant
further interpretation, but it is possible that the red wing of the
line is partially ``filled in'' by NH$_2$ emission, making the
OH$^+$(1$_2$$\to$0$_1$) line profile appear more asymmetric than the
OH$^+$(1$_1$$\to$0$_1$) line (see Section 3.3 for further
discussion). Due to the limited width of the bandpass, the
OH$^+$(1$_1$$\to$0$_1$) features are only partially covered for
HeLMS-59 and 62, but the OH$^+$(1$_2$$\to$0$_1$) lines are virtually
fully covered. As such, part of the interpretation of these sources is
based on the latter lines only.

Line parameters were obtained from Gaussian fitting to the OH$^+$ line
profiles, using two Gaussian components where appropriate. The
\ico\ lines (and other lines where appropriate) were fitted
simultaneously with additional Gaussian functions to include reliable
estimates of the uncertainties of all fitting parameters, and the
integrated line fluxes were compared to those in the moment 0 maps to
warrant internal consistency. The resulting line fluxes are reported
in Table~\ref{t2}. HeLMS-9 shows a blueshifted OH$^+$ emission
component that is sufficiently broad to be partially blended with the
\ico\ line. To reliably separate both lines, we thus compared the
\ico\ line profile to the \eco\ line profile (Fig.~\ref{f9}), which
shows a line width that is consistent with our multi-component fits to
the \ico\ and OH$^+$ features. The complexity of the OH$^+$ emission
in this source (both spatially and spectrally) warrants further
follow-up observations for a more detailed component separation.

Velocity shifts between the \ico\ (representing the systemic redshift)
and OH$^+$ absorption and emission lines from the Gaussian fitting are
provided in Table~\ref{t2b}. All shifts were calculated at the
centroid positions of each line. Values in parantheses are consistent
with no shifts at the current SNR of our data. Nine sources show no
significant velocity shifts in the absorption lines, while four and
three sources show significant red- and blueshifts of the OH$^+$
absorption, respectively. The absorption velocity shifts are in the
range of $\sim$130--360\,\kms, which is typically less than the
\ico\ line widths. Two sources show no significant velocity shifts in
the emission lines, three show significant blueshifts, and five show
significant redshifts. One source, HeLMS-9, shows components that are
both blue- and redshifted, where the latter component is closer to the
\ico\ and OH$^+$ absorption peaks. The emission velocity shifts are in
the range of $\sim$30--1000\,\kms. The largest redshifts exceed the
\ico\ line widths.

Blueshifted absorption and redshifted emission components (i.e.,
P-Cygni profiles) are expected for gas that is outflowing along the
line of sight, while the opposite (i.e., inverse P-Cygni profiles) is
expected for inflows. As such, the highest velocity emission
components may be associated with outflowing gas beyond the escape
velocities of their host galaxies. We caution that the centroid
velocities alone may not always be sufficient to distinguish between
inflows and outflows. As an example, the OH$^+$ emission centroid of
HeLMS-59 appears to be significantly redshifted relative to CO from
the spectral line fit alone (which would suggest an outflow), but it
shows an inverse P-Cygni profile, and the strongest part of the
emission is seen at blueshifted velocities. Thus, HeLMS-59 most likely
is dominated by infalling material, even though we cannot rule out the
presence of some outflowing gas. Overall, our sample appears to show
similar numbers of examples of outflows and inflows based on velocity
shifts alone under the assumption of simple, spherical geometries, and
a 3:2 ratio\footnote{This ratio would increase to 5:2 if we were to
  exclude the least significant identifications marked with a ``?'' in
  Table~\ref{t2b}.} when combining line profiles and velocity shifts,
but observations at higher spatial resolution are desirable to study
the gas kinematics and balance of outflows vs.\ inflows on scales that
come closer to resolving the starburst nuclei.

\subsubsection{Other lines}

The OH$^+$(1$_2$$\to$0$_1$) absorption feature in Orochi shows
evidence for line emission both red- and blueward of the line (see
spectra extracted centered on the peak positions of the emission in
Fig.~\ref{f8}). Since no such emission is seen near the
OH$^+$(1$_1$$\to$0$_1$) absorption feature, these components are
unlikely to be due to OH$^+$ emission, despite the fact that they
appear to be symmetrically offset from the continuum peak position
(Fig.~\ref{f5}).\footnote{We however cannot rule out the presence of
  OH$^+$ emission components on smaller scales that are diluted by the
  ACA beam with current data.}

We identify the blueward component at $\sim$221.651--221.963\,GHz
(corresponding to --830 to --400\,\kms\ relative to the systemic
redshift of OH$^+$) detected at a SNR of $\sim$3.5 with
NH(1$_2$$\to$0$_1$) line emission (which has 21 components at 974.3156
to 974.6078\,GHz, with the strongest component at 974.4784\,GHz,
redshifted to 221.9246--221.9912\,GHz and 221.9617\,GHz,
respectively). If correct, this would imply that the line centroid is
redshifted by 185$\pm$24\,\kms\ relative to the systemic velocity, at
a line FWHM of 273$\pm$61\,\kms\ (i.e., comparable to that of the
OH$^+$ absorption). Since this is significantly narrower than the
\ico\ emission line, we cannot rule out that its red wing may be
partially affected by absorption due to the (blueshifted) OH$^+$
line. This, however, is unlikely to be a strong effect due to the
similarity in shape of both detected OH$^+$ absorption lines. We thus
consider this identification to be plausible. We measure a line flux
of 1.59$\pm$0.45\,Jy\,\kms, which corresponds to a line luminosity of
$L'_{\rm NH(1-0)}$=1.1$\pm$0.3$\times$10$^{10}$\,K\,\kms\,pc$^2$.

We identify the redward component at $\sim$220.853--221.228\,GHz
(corresponding to +170 to +680\,\kms\ relative to the systemic
redshift of OH$^+$) detected at a SNR of $>$5 with
NH$_2$(3$_{12}$$\to$3$_{03}$) line emission (which has seven of its
components at 970.7079--970.7927\,GHz, with the strongest component at
970.7883\,GHz, redshifted to 221.1029--221.1222\,GHz and
221.1212\,GHz, respectively). If correct, the line centroid velocity
of 17$\pm$130\,\kms\ would be consistent with the systemic redshift,
and the line FWHM of 549$\pm$225\,\kms\ would be consistent with that
of \ico. The relatively large uncertainties on the line width are due
to the proximity to the OH$^+$ absorption feature.  We thus consider
this identification to be plausible as well.  We measure a line flux
of 3.5$\pm$1.2\,Jy\,\kms, which corresponds to a line luminosity of
$L'_{\rm NH2(312-303)}$=2.5$\pm$0.8$\times$10$^{10}$\,K\,\kms\,pc$^2$.
Higher SNR measurements are desirable to more reliably de-blend the
different emission/absorption components, to investigate any red- or
blueshifting of lines, and to more precisely constrain the
significance of any positional offsets for the NH and NH$_2$
detections at $z$=3.4 in Orochi.

As discussed in the previous section, the OH$^+$(1$_2$$\to$0$_1$)
absorption feature in HeLMS-28 appears more asymmetric and potentially
narrower than the OH$^+$(1$_1$$\to$0$_1$) absorption feature. The
shallower absorption is found at the velocities where
NH$_2$(3$_{12}$$\to$3$_{03}$) emission would be expected. While the
current SNR is too limited to attempt a reliable de-blending, we thus
consider it possible that NH$_2$ emission contributes to the apparent
OH$^+$(1$_2$$\to$0$_1$) profile. As a consequence, we base our further
interpretation of this source mainly on the OH$^+$(1$_1$$\to$0$_1$)
line.

\subsection{Position Offsets}

Given the moderate spatial resolution of our observations, the main
spatial information available are sizes and peak position offsets
between different components. In all cases where the relevant
information is available, resolved sizes are consistent with what is
expected from the Einstein radii or component separation at high
spatial resolution (Fig.~\ref{f2}), and the continuum peak positions
are consistent with those seen in the dust emission at higher
resolution. As such, we focus the remainder of our discussion on
potential offsets from the dust continuum peaks.

The peaks of the \ico\ emission coincide with those in the dust
continuum within a fraction of the synthesized beam size, with one
possible exception (Fig.~\ref{f1}). In HXMM-01, the \ico\ emission
peaks to the north-east of the dust continuum peak. HXMM-01 is the
only source in the sample that is thought to not be strongly lensed,
but it consists of two galaxies of similar brightness (and a third,
faint component) in the dust continuum (Fig.~\ref{f2}; see also Fu et
al.\ \citeyear{fu13}; Bussmann et al.\ \citeyear{bussmann15}). The
peak of the \ico\ emission appears consistent with the northern of the
two galaxies. Our observations thus appear to suggest that this galaxy
dominates the high-$J$ line emission.

The peaks of all OH$^+$ absorption lines coincide with the dust
continuum peaks in all cases, as expected. Minor offsets are only seen
in the sources with the lowest SNRs, and thus, are not significant
given the noise variations in the maps.

The OH$^+$ emission peaks are coincident with the dust continuum peaks
in most cases, with a few possible exceptions. The emission in
HeLMS-65 is too weak for the offset to be considered reliable. HeLMS-9
shows two emission components, one for each of the two lensed
arcs. The redshifted component is superposed in velocity with the
OH$^+$ absorption component, such that the apparent offset from the
dust continuum peak may be due to the fact that the emission is
reduced due to absorption in the foreground at that position. The
blueshifted peak is lower SNR, and it also is impacted by a mix of
absorption and emission near the systemic redshift.  HeLMS-45 (which
appears to show two emission components) and 44 (which may show a
tentative offset in a similar direction as CO), may also show spatial
offsets, but in contrast to HeLMS-9, it remains unclear that there is
a significant velocity overlap between the absorption and emission
components.  Since the SNR of the emission components for both of
these sources is only moderate, it cannot be ruled out with the
current data that the spatial offsets are due to residual
uncertainties in the de-blending of different components. Higher
resolution imaging at better sensitivity is required to provide a more
detailed interpretation for any of these sources. Comparing the
emission components in both OH$^+$ lines for HeLMS-59, the apparent
finite extent seen in the OH$^+$(1$_1$$\to$0$_1$) line may be due to the
less reliable phase calibration at the outer edge of the bandpass,
where the emission peaks. We thus do not consider it significant, and
we focus on the (fully covered) OH$^+$(1$_2$$\to$0$_1$) line for most of
the interpretation in the following.

For J0210+0016, the OH$^+$ emission appears to peak closer to the
center of the main lensed arc of the source than the dust
continuum. This source is thought to contain an AGN (e.g., Geach et
al.\ \citeyear{geach18}), which may impact the OH$^+$ formation
mechanisms in part of the galaxy. Since J0210+0016 shows no evidence
for OH$^+$ absorption at the sensitivity limit of our data, we
conclude that it is the only target for which the OH$^+$ emission peak
offset can be considered significant at present.

\begin{figure*}%[tbh]
\epsscale{0.605}
\plotone{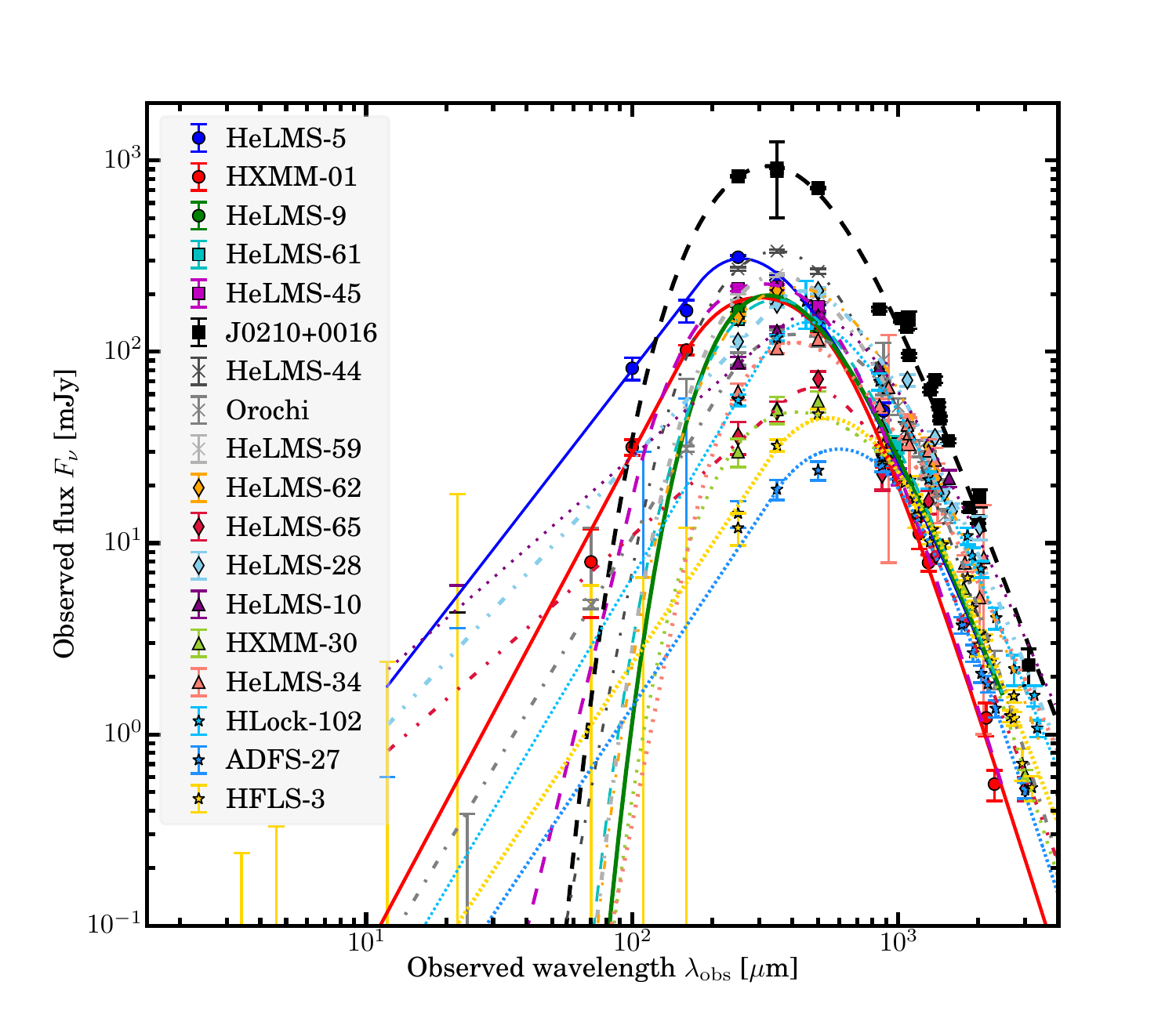}
\epsscale{0.53}
\plotone{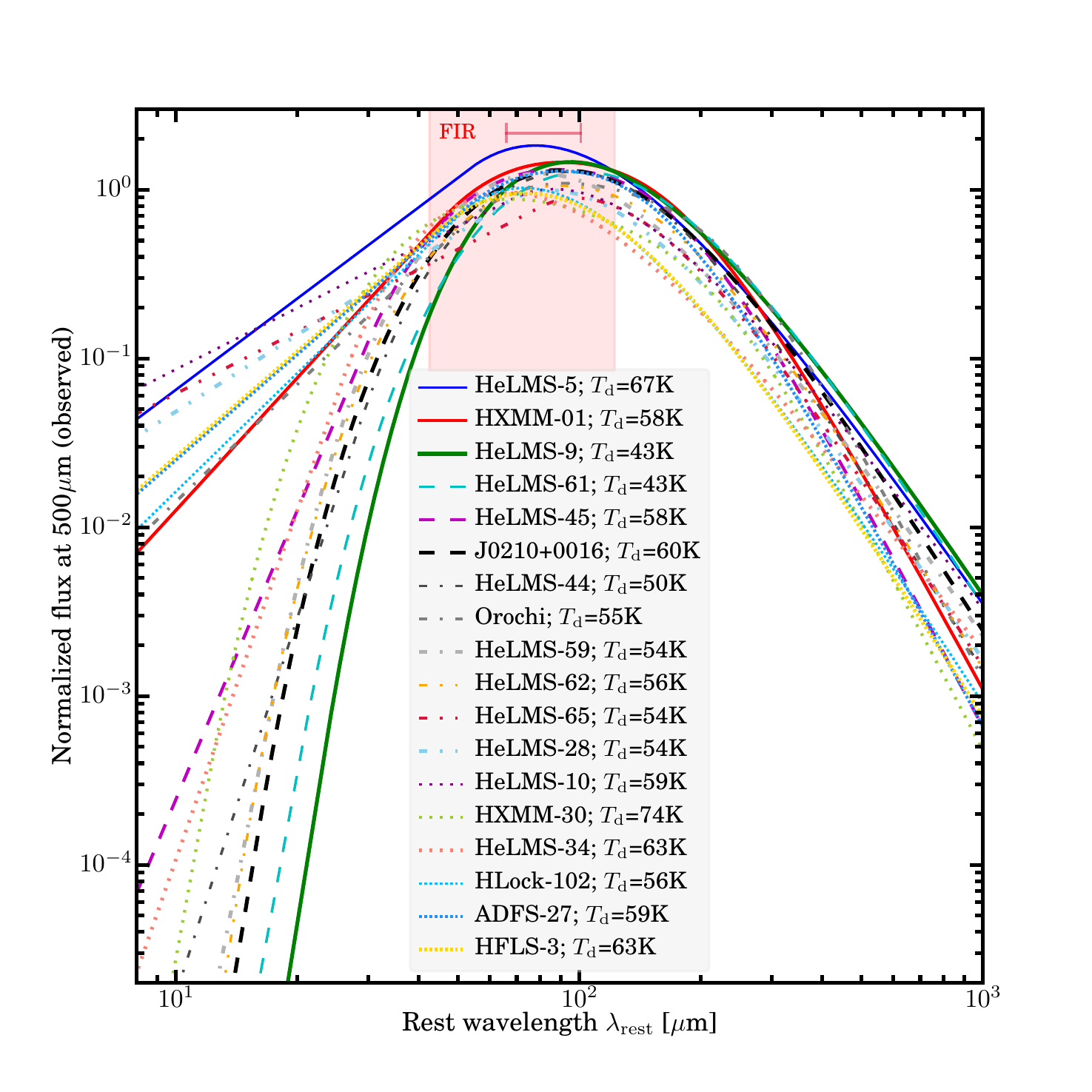}
\vspace{-5mm}

\caption{Spectral energy distributions of the GADOT sample and MBB SED fits in the observed frame (left; data points and lines), and models in the rest-frame, normalized to the observed 500\,$\mu$m flux (right), including ADFS-27 and HFLS3 (Riechers et al.\ \citeyear{riechers13b,riechers17,riechers20c}). The shaded region indicates the wavelength range used to calculate the far-infrared luminosities, while the entire range shown is used to calculate total infrared luminosities. The bar at the top indicates the range in peak wavelengths.\label{f10}}
%\vspace{-5mm}
%
\end{figure*}

\section{Analysis and Discussion}

\subsection{Spectral Energy Distribution Modeling}

To determine the dust properties of our sample, we fit modified
black-body models to the rest-frame far-infrared to submillimeter
wavelength emission using the Markov Chain Monte Carlo-based approach
{\sc mbb\_emcee} described in more detail by Riechers et
al.\ (\citeyear{riechers13b}) and Dowell et
al.\ (\citeyear{dowell14}). This method uses the dust temperature
$T_{\rm dust}$, the spectral slope of the dust emissivity $\beta_{\rm
  IR}$, and the wavelength $\lambda_0$ where the optical depth reaches
unity as the main fit parameters. To capture the approximate shape of
the spectral energy distribution (SED) on the far Wien side, we join
the MBB to a $\nu^{\alpha}$ power law at short wavelengths. The
observed-frame 500\,$\mu$m flux density is used as a normalization
parameter of the fits. To guide the fitting process, we place a broad
prior of $\beta_{\rm IR}$=1.8$\pm$0.6 on the emissivity parameter
(i.e., centered on the value for Milky Way molecular clouds; e.g.,
Planck Collaboration \citeyear{planck11}), and leave all other
parameters without a prior. The results for the entire sample are
reported in Table~\ref{t3} and shown in Fig.~\ref{f10}, also including
our previous fits for ADFS-27 and HFLS3 (Riechers et
al.\ \citeyear{riechers13b,riechers20c}). We also re-fit four
comparison sources at $z$=2.3--3.0 with either OH$^+$ or
\ico\ detections (including two sources where both lines were either
solidly or tentatively detected) from the literature to enable a
meaningful comparison, and have included the resulting values in
Table~\ref{t3}.

For the GADOT sample, we find median values of $T_{\rm
  dust}$=56.9$\pm$3.1\,K, $\beta_{\rm IR}$=2.0$\pm$0.3, and
$\lambda_0$=201$\pm$38\,$\mu$m, where quoted uncertainties are the
median absolute deviation. The median peak wavelength of the SEDs is
$\lambda_{\rm peak}$=86$\pm$7\,$\mu$m. The range in apparent $S_{250}$
is a factor of $\sim$70 ($\sim$30 when only including strongly-lensed
sources), while the range in apparent far-infrared luminosity $L_{\rm
  FIR}$ is only a factor of $\sim$8. Since the range in $T_{\rm dust}$
($\lambda_{\rm peak}$) is only a factor of $\sim$1.7 (1.5), this
difference is largely due to the large range in redshift spanned by
the sample. This is consistent with the factor of $\sim$7 range in
$S_{870}$, i.e., at a wavelength where the negative $K$-correction is
much stronger than at 250\,$\mu$m. The range in $\lambda_0$ is a
factor of $\sim$2.9, but for all sources, $\lambda_0$ is smaller than
the rest wavelengths of the OH$^+$ 1$_1$$\to$0$_1$ and 1$_2$$\to$0$_1$
lines. Thus, the dust is expected to only be moderately optically
thick towards (or behind) the observed absorption/emission lines.

%\begin{turnpage}

\begin{figure*}
\begin{deluxetable*}{ l c c c c c c c c c c c c c c c }

%\tabletypesize{\scriptsize}
\tablecaption{{\em Herschel}/SPIRE and 870\,$\mu$m fluxes, \ico\ luminosities, and parameters obtained from dust spectral energy distribution fitting to our sample and four sources in the literature (including two secondary sources from the master sample). \label{t3}}
\tablehead{
  Name     & redshift & $S_{250}$ & $S_{350}$ & $S_{500}$ & $S_{870}$\tablenotemark{a} & $T_{\rm dust}$ & $\beta_{\rm IR}$ & $\lambda_{\rm peak}$ & $\lambda_0$ & $L_{\rm FIR}$\tablenotemark{b,c} & $L_{\rm IR}$\tablenotemark{b,c,d} & $L'_{\rm CO(9-8)}$\tablenotemark{b,e} & $L_{\rm CO(9-8)}$\tablenotemark{b} & References\\
           &          & (mJy)    & (mJy) & (mJy)      & (mJy) & (K)           &                & ($\mu$m)           & ($\mu$m)   & (10$^{12}$\,\lsol ) & (10$^{12}$\,\lsol ) & (10$^{10}$\,$L_l$ ) & (10$^8$\,\lsol ) & }
\startdata
HeLMS-5    & 2.2919 & 312$\pm$6 & 244$\pm$7 & 168$\pm$8 & 49$\pm$5  & 66.5$^{+5.1}_{-6.8}$   & 2.05$^{+0.48}_{-0.45}$ & 74$^{+4}_{-3}$  & 205$^{+22}_{-52}$ & 40.0$^{+1.4}_{-1.6}$  & 68.1$^{+6.7}_{-5.7}$  & 2.92$\pm$0.45 & 10.4$\pm$1.6 & 1 \\
HXMM-01    & 2.3079 & 180$\pm$7 & 192$\pm$8 & 132$\pm$7 & 29$\pm$3  & 57.8$^{+1.4}_{-2.8}$   & 2.52$^{+0.18}_{-0.11}$ & 88$^{+4}_{-2}$  & 236$^{+11}_{-37}$ & 22.8$^{+1.0}_{-0.4}$  & 37.3$^{+1.8}_{-1.5}$  & 2.37$\pm$0.39 & 8.5$\pm$1.4 & 1 \\
HeLMS-9    & 2.3934 & 166$\pm$6 & 195$\pm$6 & 135$\pm$7 &           & 43.0$^{+7.3}_{-8.6}$   & 1.82$^{+0.23}_{-0.23}$ & 95$^{+3}_{-2}$  & 138$^{+58}_{-89}$ & 21.2$^{+1.5}_{-1.6}$  & 33.4$^{+1.1}_{-7.0}$  & 2.32$\pm$0.45 & 8.3$\pm$1.6 & 1 \\
HeLMS-61   & 2.4022 & 148$\pm$6 & 187$\pm$6 & 147$\pm$9 &           & 42.7$^{+5.4}_{-7.3}$   & 1.91$^{+0.32}_{-0.31}$ & 101$^{+4}_{-3}$ & 170$^{+51}_{-87}$ & 19.3$^{+1.3}_{-1.4}$  & 30.3$^{+0.7}_{-5.7}$  & 1.33$\pm$0.17 & 4.73$\pm$0.60 & 1 \\
HeLMS-45   & 2.4726 & 214$\pm$7 & 218$\pm$7 & 172$\pm$9 &           & 58.0$^{+2.7}_{-3.6}$   & 2.70$^{+0.38}_{-0.35}$ & 87$^{+4}_{-3}$  & 208$^{+8}_{-24}$  & 31.2$^{+1.3}_{-1.5}$  & 46.3$^{+2.8}_{-4.5}$ & 6.92$\pm$0.99 & 24.7$\pm$3.5 & 1 \\
J0210+0016 & 2.5534 & 826$\pm$7 & 912$\pm$7 & 717$\pm$8 & 167$\pm$4 & 59.8$^{+6.3}_{-8.6}$   & 2.01$^{+0.07}_{-0.14}$ & 87$^{+5}_{-3}$  & 256$^{+63}_{-80}$ & 130.6$^{+3.4}_{-6.8}$ & 262.0$^{+70.0}_{-91.5}$ & 7.43$\pm$0.66 & 26.5$\pm$2.4 & 1 \\
HeLMS-44   & 2.6895 & 271$\pm$6 & 336$\pm$6 & 263$\pm$8 &           & 50.3$^{+2.2}_{-4.2}$   & 2.15$^{+0.28}_{-0.26}$ & 93$^{+3}_{-1}$  & 197$^{+3}_{-51}$  & 47.7$^{+1.5}_{-2.0}$  & 73.7$^{+0.2}_{-11.6}$ & 4.86$\pm$0.80 & 17.3$\pm$2.9 & 1 \\
Orochi     & 3.3903 &  92$\pm$7 & 122$\pm$8 & 113$\pm$7 & 63$\pm$6  & 55.0$^{+1.3}_{-1.5}$   & 2.75$^{+0.26}_{-0.27}$ & 93$^{+3}_{-2}$  & 274$^{+11}_{-16}$ & 27.4$^{+1.0}_{-0.8}$  & 48.7$^{+1.3}_{-1.1}$ & 2.50$\pm$0.38 & 8.9$\pm$1.4 & 1 \\
HeLMS-59   & 3.4346 & 193$\pm$7 & 252$\pm$6 & 202$\pm$8 &           & 53.7$^{+6.6}_{-9.4}$   & 1.69$^{+0.18}_{-0.20}$ & 79$^{+2}_{-1}$  & 127$^{+50}_{-77}$ & 60.1$^{+1.6}_{-1.6}$  & 92.5$^{+5.0}_{-14.2}$ & 5.65$\pm$0.63 & 20.1$\pm$2.3 & 1 \\
HeLMS-62   & 3.5027 & 151$\pm$6 & 209$\pm$6 & 205$\pm$8 &           & 56.1$^{+2.0}_{-2.1}$   & 2.21$^{+0.44}_{-0.43}$ & 88$^{+3}_{-2}$  & 209$^{+22}_{-26}$ & 52.1$^{+1.2}_{-1.2}$  & 76.8$^{+2.1}_{-4.8}$  & 7.7$\pm$1.5 & 27.6$\pm$5.4 & 1 \\
HeLMS-65   & 3.7966 &  36$\pm$7 &  49$\pm$6 &  72$\pm$7 & 23$\pm$4  & 53.6$^{+7.7}_{-7.9}$   & 2.10$^{+0.21}_{-0.21}$ & 90$^{+7}_{-7}$  & 205$^{+50}_{-51}$ & 15.7$^{+1.4}_{-1.4}$  & 24.2$^{+3.1}_{-4.1}$  & 1.20$\pm$0.28 & 4.3$\pm$1.0 & 1 \\
HeLMS-28   & 4.1625 & 113$\pm$7 & 177$\pm$6 & 209$\pm$8 & 82$\pm$4  & 53.8$^{+5.0}_{-6.8}$   & 1.74$^{+0.16}_{-0.17}$ & 82$^{+2}_{-2}$  & 150$^{+24}_{-48}$ & 63.5$^{+1.7}_{-1.6}$  & 97.3$^{+18.5}_{-12.2}$ & 7.21$\pm$0.60 & 25.7$\pm$2.2 & 1 \\
HeLMS-10   & 4.3726 &  88$\pm$6 & 129$\pm$6 & 155$\pm$7 & 82$\pm$5  & 58.8$^{+3.8}_{-3.2}$   & 1.66$^{+0.32}_{-0.33}$ & 82$^{+3}_{-3}$  & 208$^{+40}_{-39}$ & 53.7$^{+1.6}_{-1.6}$  & 88.0$^{+14.8}_{-12.0}$ & 4.39$\pm$0.78 & 15.7$\pm$2.8 & 1 \\
HXMM-30    & 5.094  &  30$\pm$5 &  50$\pm$8 &  55$\pm$7 & 28$\pm$2  & 74.2$^{+4.7}_{-4.6}$   & 2.64$^{+0.37}_{-0.39}$ & 69$^{+4}_{-4}$  & 242$^{+29}_{-34}$ & 24.7$^{+1.8}_{-1.7}$  & 44.1$^{+4.2}_{-6.4}$ & 2.56$\pm$0.52 & 9.2$\pm$1.9 & 1 \\
HeLMS-34   & 5.1614 &  62$\pm$6 & 104$\pm$6 & 116$\pm$7 & 41$\pm$3  & 63.0$^{+7.8}_{-8.3}$   & 1.36$^{+0.14}_{-0.18}$ & 66$^{+2}_{-2}$  & 93$^{+50}_{-64}$  & 53.8$^{+2.1}_{-2.0}$  & 86.5$^{+5.1}_{-7.8}$ & 6.98$\pm$0.67 & 24.9$\pm$2.4 & 1 \\
HLock-102  & 5.2915 &  56$\pm$4 & 118$\pm$4 & 139$\pm$4 & 70$\pm$7  & 55.7$^{+5.1}_{-7.5}$   & 1.80$^{+0.07}_{-0.08}$ & 73$^{+2}_{-1}$  & 99$^{+31}_{-43}$  & 68.2$^{+1.5}_{-1.3}$  & 96.7$^{+3.4}_{-5.0}$ & 12.1$\pm$1.9 & 43.1$\pm$6.7 & 1 \\
ADFS-27    & 5.6550 &  14$\pm$2 &  19$\pm$2 &  24$\pm$2 & 25$\pm$2  & 59.2$^{+3.3}_{-4.1}$   & 2.52$^{+0.19}_{-0.17}$ & 85$^{+5}_{-4}$  & 191$^{+11}_{-19}$ & 15.8$^{+1.0}_{-1.9}$ & 23.8$^{+2.3}_{-2.2}$ & 2.48$\pm$0.06 & 8.84$\pm$0.20 & 2, 3 \\
HFLS3      & 6.3369 &  12$\pm$2 &  32$\pm$2 &  47$\pm$3 & 33$\pm$2  & 63.3$^{+5.4}_{-5.8}$   & 1.94$^{+0.07}_{-0.09}$ & 73$^{+2}_{-1}$  & 142$^{+25}_{-27}$ & 29.3$^{+1.4}_{-1.3}$ & 55.0$^{+3.0}_{-2.2}$ & 4.51$\pm$0.73 & 16.1$\pm$2.6 & 4, 5 \\
Literature: & & & & & & & & & & & & & & \\
SDP-17b\tablenotemark{f} & 2.3051 & 354$\pm$7 & 339$\pm$8 & 220$\pm$9 & 55$\pm$3  & 60.3$^{+2.8}_{-4.8}$   & 2.32$^{+0.25}_{-0.24}$ & 83$^{+4}_{-3}$  & 204$^{+13}_{-27}$ & 45.3$^{+1.7}_{-2.2}$ & 70.9$^{+3.6}_{-9.5}$ & & & 6 \\
Eyelash    & 2.3259 & 366$\pm$55 & 429$\pm$64 & 325$\pm$49 & 106$\pm$12 & 48.3$^{+1.1}_{-2.0}$ & 2.81$^{+0.18}_{-0.17}$ & 105$^{+3}_{-2}$ & 223$^{+10}_{-25}$ & 44.9$^{+2.9}_{-1.9}$ & 66.2$^{+4.1}_{-2.8}$ & 1.19$\pm$0.56 & 4.2$\pm$2.0 & 7, 8 \\
HerBS-89a  & 2.9497 &  72$\pm$6 & 103$\pm$6 &  96$\pm$7 & 53$\pm$4  & 49.7$^{+2.6}_{-2.8}$   & 2.27$^{+0.10}_{-0.10}$ & 102$^{+5}_{-5}$  & 256$^{+20}_{-22}$ & 16.8$^{+1.0}_{-1.0}$ & 26.0$^{+0.7}_{-3.1}$ & 0.88$\pm$0.10 & 3.14$\pm$0.37 & 9 \\
HLock-01\tablenotemark{f} & 2.9574 & 425$\pm$10 & 340$\pm$10 & 233$\pm$11 & 53$\pm$1 & 77.4$^{+2.0}_{-3.0}$ & 1.93$^{+0.17}_{-0.19}$ &  64$^{+1}_{-1}$  & 151$^{+15}_{-20}$ & 87.1$^{+1.9}_{-1.1}$ & 176.7$^{+4.7}_{-2.8}$ & 6.3$\pm$1.6 & 22.5$\pm$5.8 & 10, 11, 12 \\
%\vspace{-2mm}
\enddata
\tablenotetext{\rm a}{Fluxes for J0210+0016, HeLMS-28, 10, HLock-102, and HerBS-89a are measured at 850\,$\mu$m. Fluxes for HeLMS-5, HFLS3, SDP-17b, the Eyelash, and HLock-01 are measured at 880\,$\mu$m.}
\tablenotetext{\rm b}{Apparent values not corrected for gravitational magnification where applicable. Published lens magnification factors are:\ HeLMS-5 $\mu_{\rm L}$=8.3$\pm$0.6 (Dye et al.\ \citeyear{dye18}); HXMM-01 subcomponents:\ $\mu_{\rm L}$=1.39$\pm$0.19; 1.21$\pm$0.01; 1.29$\pm$0.15 (Bussmann et al.\ \citeyear{bussmann15}); J0210+0016 $\mu_{\rm L}$=14.7$\pm$0.3 (Geach et al.\ \citeyear{geach18}); Orochi $\mu_{\rm L}$=5.33$\pm$0.19 (Bussmann et al.\ \citeyear{bussmann15}); HLock-102 $\mu_{\rm L}$=12.5$\pm$1.2 (Riechers et al.\ \citeyear{riechers20a}); HFLS3 $\mu_{\rm L}$=1.8$\pm$0.6 (Riechers et al.\ \citeyear{riechers20a}); SDP17b $\mu_{\rm L}$=4.9$\pm$0.7 (Bussmann et al.\ \citeyear{bussmann13}); Eyelash $\mu_{\rm L}$=37.5$\pm$4.5 (Swinbank et al.\ \citeyear{swinbank11}); HerBS-89a $\mu_{\rm L}$=5.05$\pm$0.03 (Berta et al.\ \citeyear{berta20}); HLock-01 $\mu_{\rm L}$=10.9$\pm$0.7 (Gavazzi et al.\ \citeyear{gavazzi11}).}
\tablenotetext{\rm c}{$L_{\rm FIR}$ ($L_{\rm IR}$) is integrated over the 42.5--122.5\,$\mu$m (8--1000\,$\mu$m) range in the rest frame.}
\tablenotetext{\rm d}{Multiply $L_{\rm IR}$ by a factor of 1.0$\times$10$^{-10}$ to obtain apparent star formation rates (SFRs), under the assumption of a Chabrier (\citeyear{chabrier03}) stellar initial mass function.}
\tablenotetext{\rm e}{Given in units of $L_l$=K\,\kms\,pc$^2$.}
\tablenotetext{\rm f}{Source is also part of the GADOT master sample, but not re-analyzed with the same methods as used for the sample studied here.}
\tablereferences{[1] this work; [2--5] Riechers et al.\ (\citeyear{riechers20c,riechers17,riechers20a,riechers13b}); [6] Bussmann et al.\ (\citeyear{bussmann13}); [7] Swinbank et al.\ (\citeyear{swinbank10}); [8] Danielson et al.\ (\citeyear{danielson11}); [9] Berta et al.\ \citeyear{berta20}; [10] Conley et al.\ (\citeyear{conley11}); [11] Riechers et al.\ (\citeyear{riechers11d}); [12] Scott et al.\ (\citeyear{scott11}).}
\vspace{-9mm}
\end{deluxetable*}
%\vspace{-9mm}

\end{figure*}

%\clearpage
%\end{turnpage}

\begin{figure*}%[tbh]
\epsscale{0.565}
\plotone{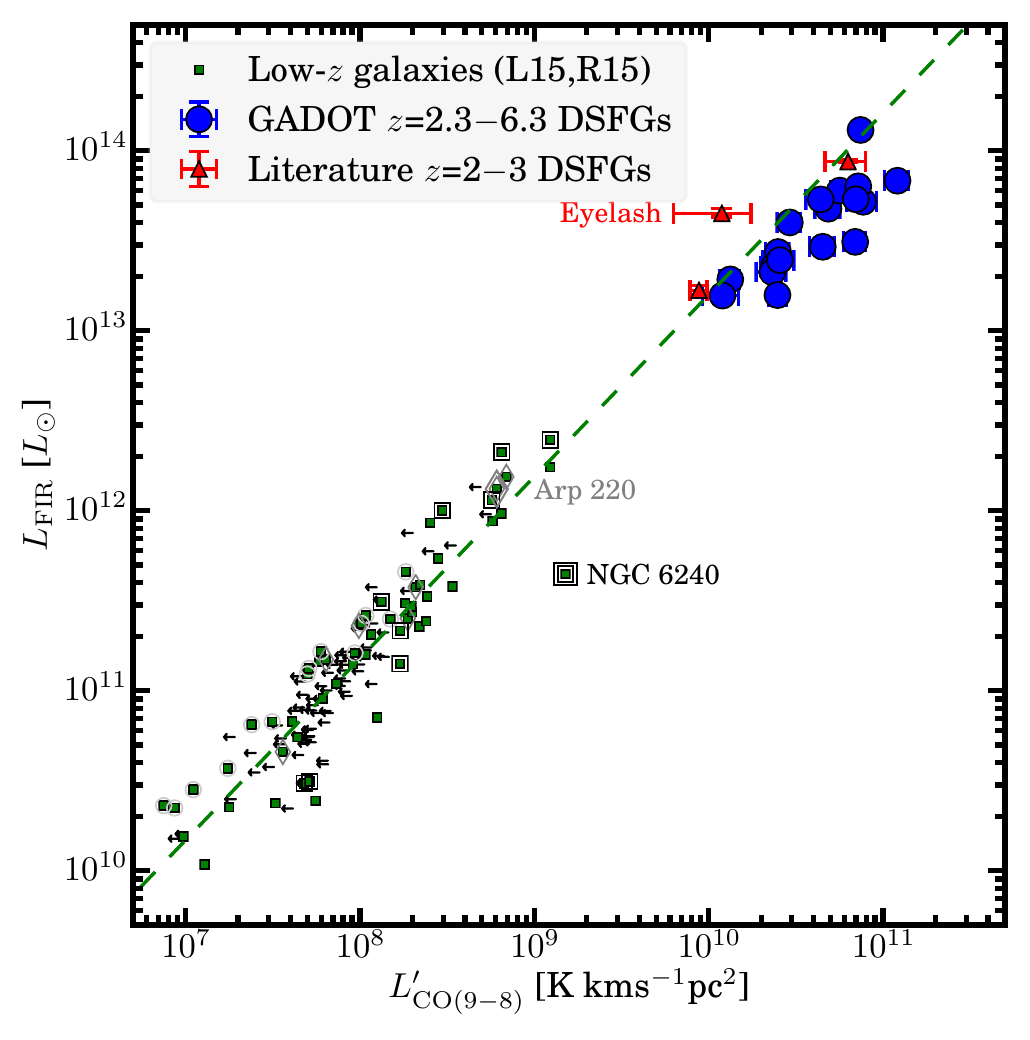}
\epsscale{0.585}
\plotone{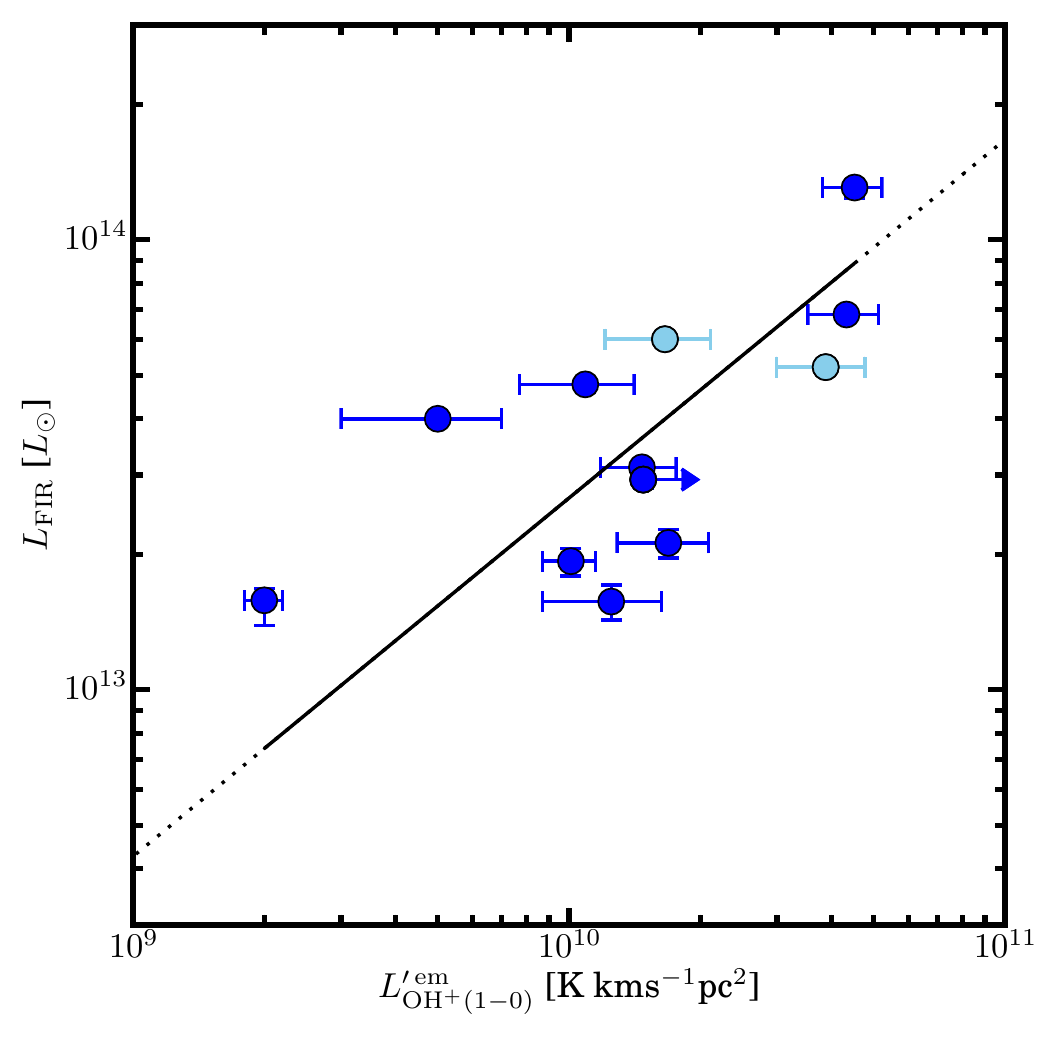}
%\vspace{-5mm}

\caption{Left:\ \ico\ -- FIR luminosity relation for the GADOT sample (blue dots), other {\em Herschel}-selected DSFGs, and the Cosmic Eyelash (red triangles; see Table~\ref{t3}) and nearby galaxies (green symbols and upper limit arrows; Liu et al.\ \citeyear{liu15}; L15). The dashed line is the fit to the nearby galaxy sample proposed by Liu et al.\ (\citeyear{liu15}), which has a power-law slope of $N$=1.01. Fluxes were not corrected for gravitational lensing magnification. Given the linear slope of the relation, lensing effects are unimportant unless the dust and warm molecular gas are differentially lensed relative to each other. Gray circle, diamond, and square outlines indicate Class I, II, and III sources from the HerCULES survey as identified by Rosenberg et al.\ (\citeyear{rosenberg15}; R15), respectively. Higher classes indicate higher CO excitation. The GADOT sample lies between the luminosity ratios of Arp\,220 and NGC\,6240. The systematic offset of the sample towards higher $L^\prime_{\rm CO(9-8)}$ may indicate a more important role of shock heating compared to the nearby star-forming galaxies. Right:\ OH$^+$ emission -- FIR luminosity relation (solid line) for the GADOT sample. Light blue dots indicate the sources for which OH$^+$(1$_2$$\to$0$_1$) is used instead of OH$^+$(1$_1$$\to$0$_1$) since the emission components in the latter were not fully covered.\label{f12}}
%\vspace{-5mm}
%
\end{figure*}

\subsection{Scaling Relations}

\subsubsection{\ico\ -- FIR and OH$^+$ -- FIR Relations}

We find a strong correlation between the \ico\ line ($L^\prime_{\rm
  CO(9-8)}$) and FIR dust continuum luminosities for the GADOT sample
(Fig.~\ref{f12} left). This relation is reminiscent for that found for
nearby star-forming galaxies (e.g., Liu et al. \citeyear{liu15}),
consistent with the idea that the \ico\ emission is associated with
warm, dense molecular gas in the star-forming regions. The GADOT
sample is consistent with the linear slope of the relation found for
the lower-luminosity nearby galaxies, but the sample appears
systematically offset toward higher \ico\ luminosity.

At first glance, one may conclude that this could be due to the fact
that most GADOT sources are strongly lensed, given that no
magnification correction has been applied. However, given the linear
slope of the relation, such corrections would only be important if
there is {\em differential} lensing between the $\sim$70--100\,$\mu$m
dust near the peak of the SEDs and the warm, dense molecular
gas. Also, the observed offset is in the opposite direction of what
would be expected for differential lensing between a compact warm dust
component and a more extended molecular gas component.

Another possibility would be that the gas excitation in the GADOT
sample is systematically higher than in the nearby galaxies. To
further investigate this issue, Fig.~\ref{f12} highlights the galaxies
that overlap with the Class I, II, and III samples of the HerCULES
survey (Rosenberg et al.\ \citeyear{rosenberg15}), where a higher
class corresponds to a higher CO excitation. Indeed, Class I sources
appear to show systematically lower \ico\ to FIR luminosity ratios
than the Class II and III samples, and Class II sources appear to
preferentially fall within the intermediate range. This is consistent
with the idea that differences in CO excitation may be responsible for
a significant portion of the observed scatter of the relation. Class
III sources show the broadest scatter, with about half of them being
below the relation, in a similar region as the GADOT sources. The
strongest outlier is the binary active galactic nucleus NGC\,6240,
which shows the highest $L^\prime_{\rm CO(9-8)}$/$L_{\rm FIR}$ ratio
among the entire combined low- and high-redshift sample. The high CO
line to dust continuum luminosity ratio in NGC\,6240 is thought to be
due to shock excitation (e.g., Meijerink et
al.\ \citeyear{meijerink13}). Thus, at face value, our findings may
suggest more prevalent shock excitation (rather than stronger
radiative excitation due to the warm dust or hidden active galactic
nuclei) in DSFGs at $z$=2--6 than in most nearby star-forming
galaxies, perhaps due to slow-moving shocks. A possible alternative
explanation would be that these galaxies exhibit increased cosmic ray
energy densities which lead to an increased CO excitation.\footnote{In
  principle, heating from hard radiation sources such as X-rays cannot
  be ruled out based on the current data alone, even if it is not our
  preferred explanation due to the lack of other evidence for the
  presence of hard X-ray sources in these galaxies.}

We also find a weak correlation between the OH$^+$ line emission
components ($L^{\prime\,{\rm em}}_{\rm OH^+(1-0)}$) and FIR dust
continuum luminosities for the GADOT sample (Fig.~\ref{f12} right),
with a best fit of log($L_{\rm FIR}$/\lsol
)=(5.5$\pm$2.5)+(0.35$\pm$0.10)\,log($L^{\prime\,{\rm em}}_{\rm
  OH^+(1-0)}$/$L_l$), where $L_l$=K\,\kms\,pc$^2$. The OH$^+$ emission
components are consistent with being associated with shock-heated,
dense gas components impacted by galactic winds (e.g., Falgarone et
al.\ \citeyear{falgarone17}; Riechers et
al.\ \citeyear{riechers20c}). As discussed above, the presence of
shock-heated gas is consistent with the strong \ico\ emission across
the GADOT sample. Thus, while we cannot make strong inferences based
on this relation alone, and while alternative scenarios cannot be
ruled out, our findings for the CO and OH$^+$ emission paint a
consistent picture for the heating and excitation mechanisms in
$z$=2--6 DSFGs.

\begin{figure*}[tbh]
\epsscale{0.36}
\plotone{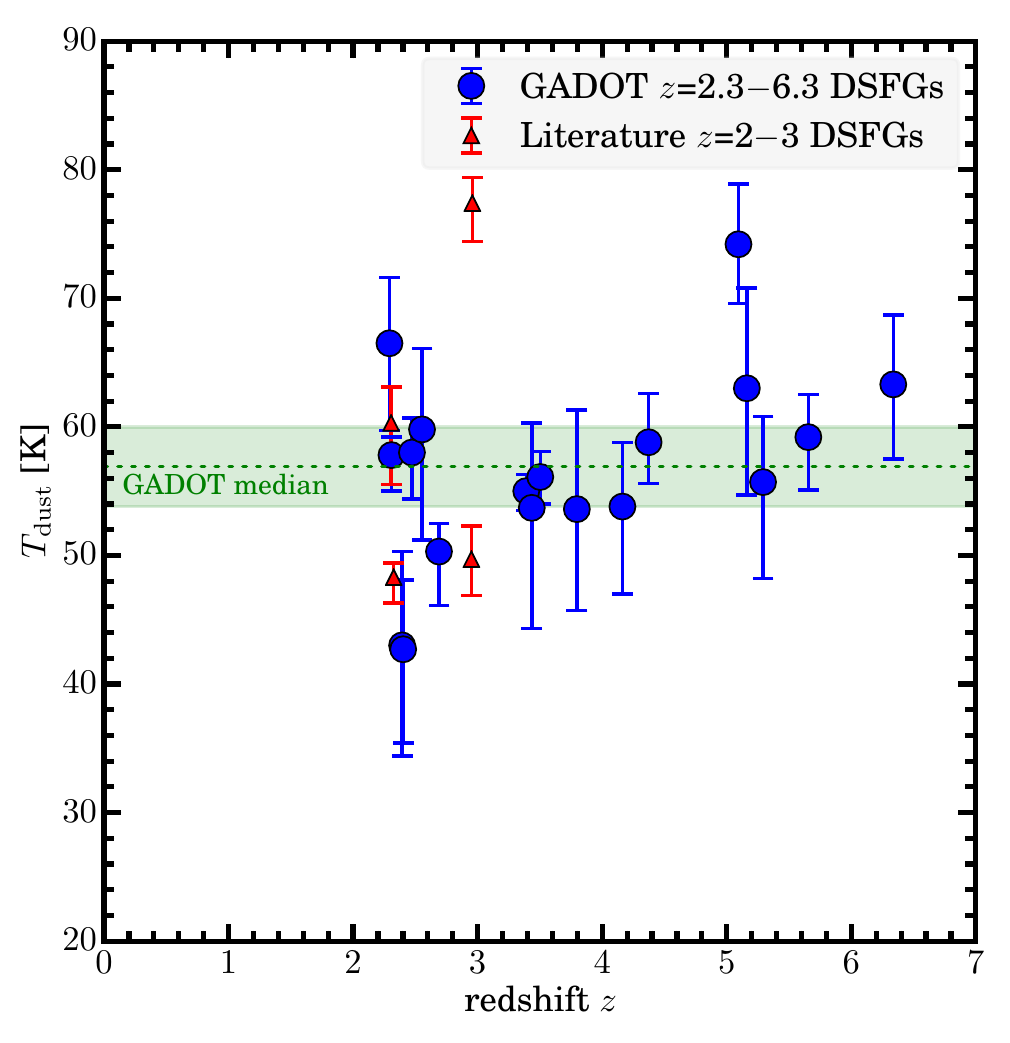}
\plotone{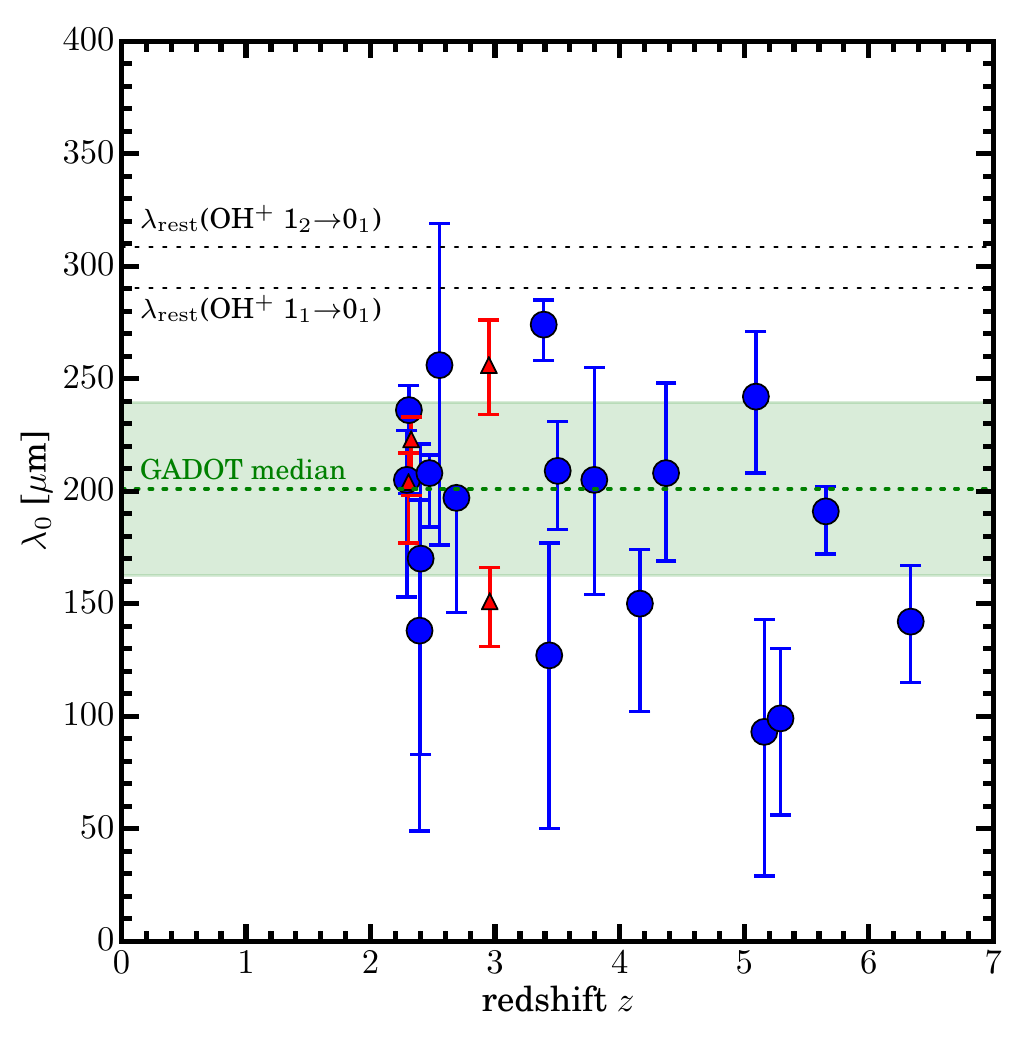}
\plotone{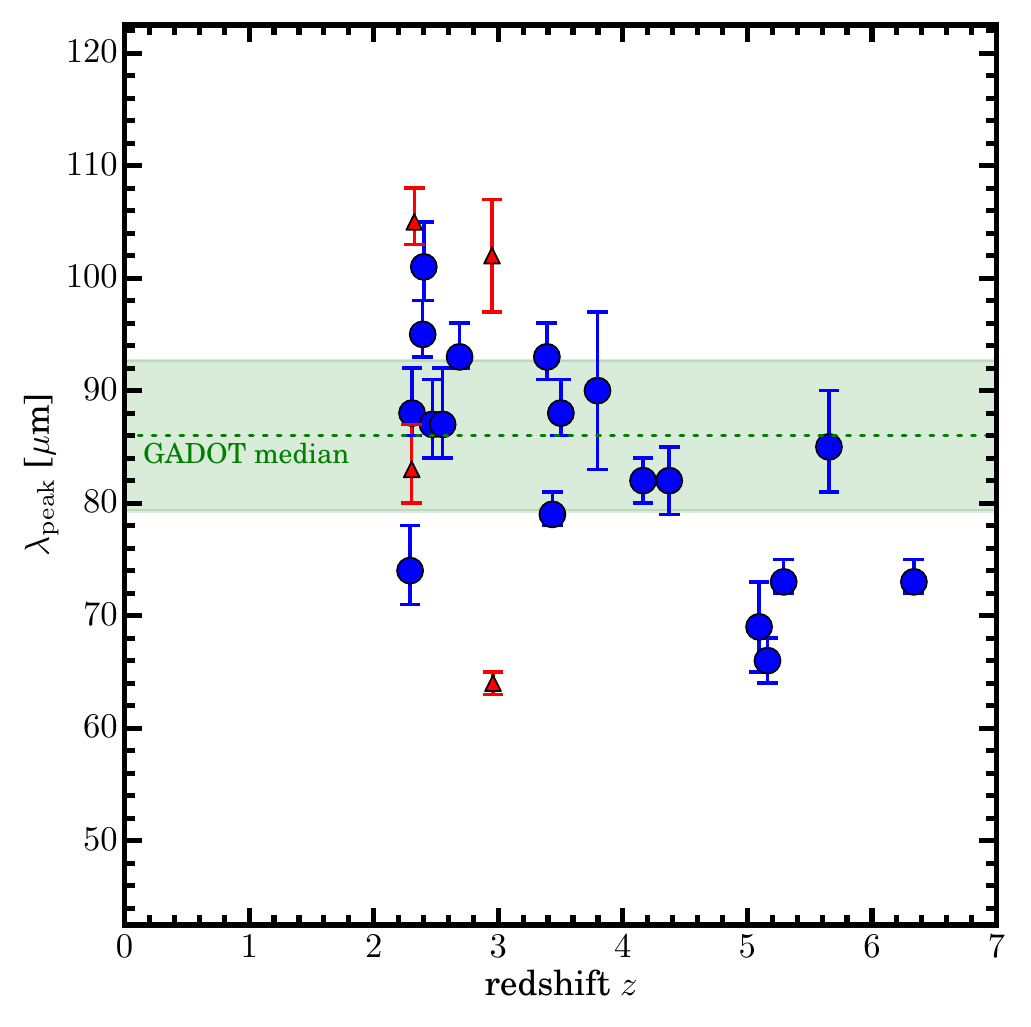}
\plotone{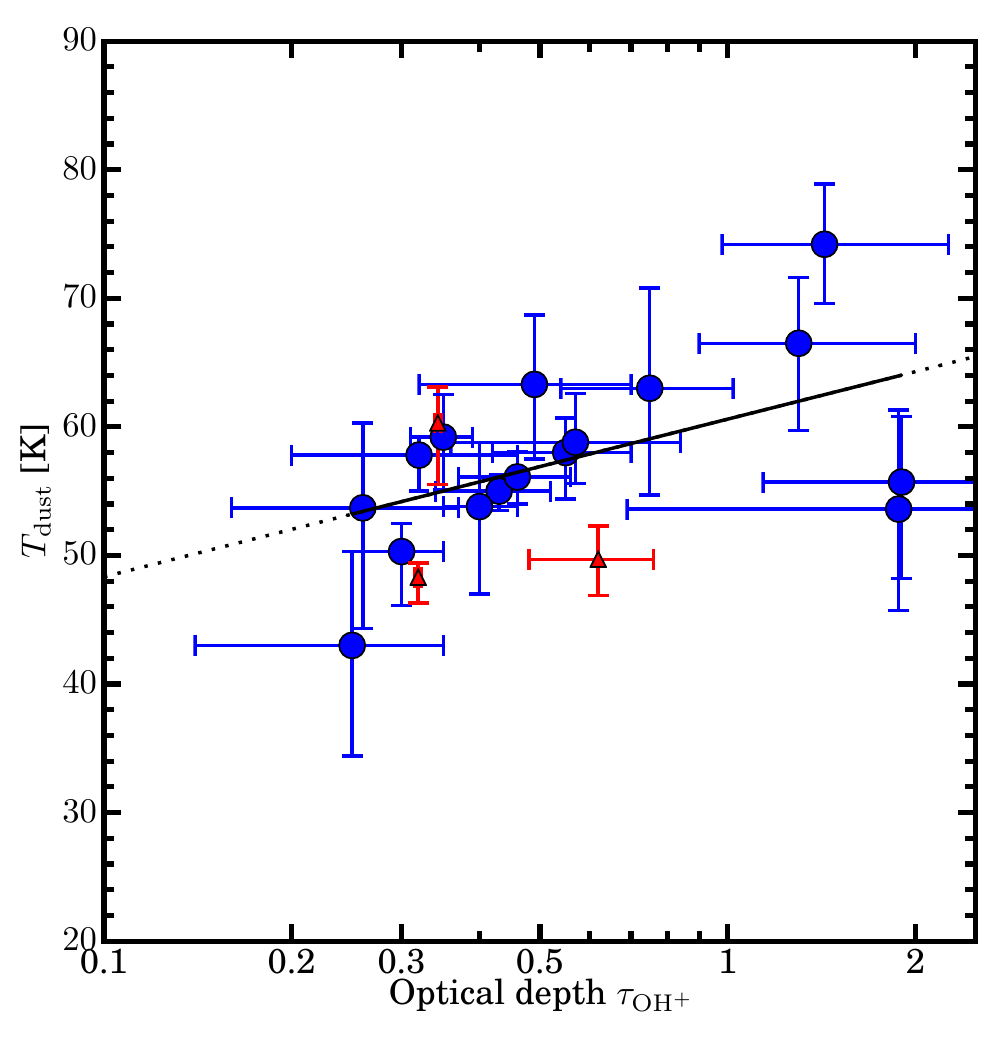}
\plotone{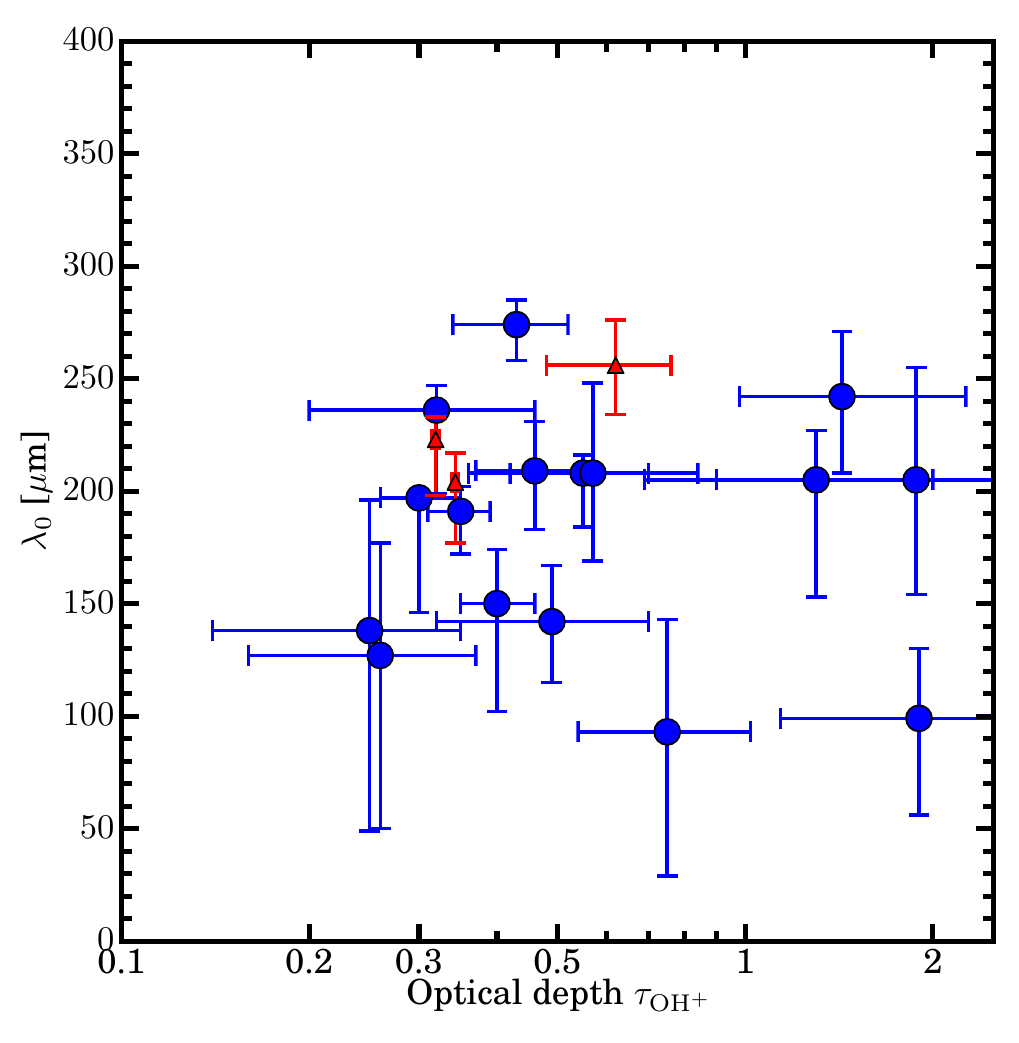}
\plotone{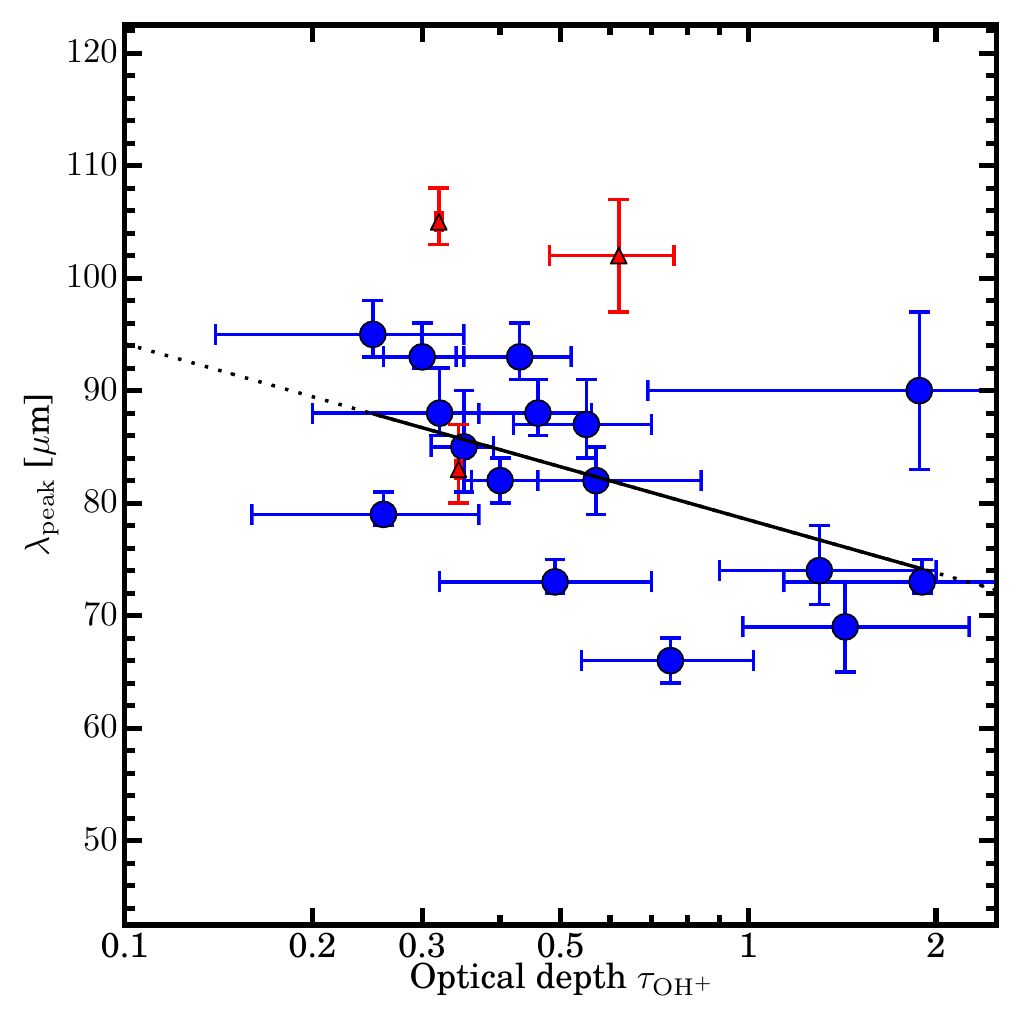}
\plotone{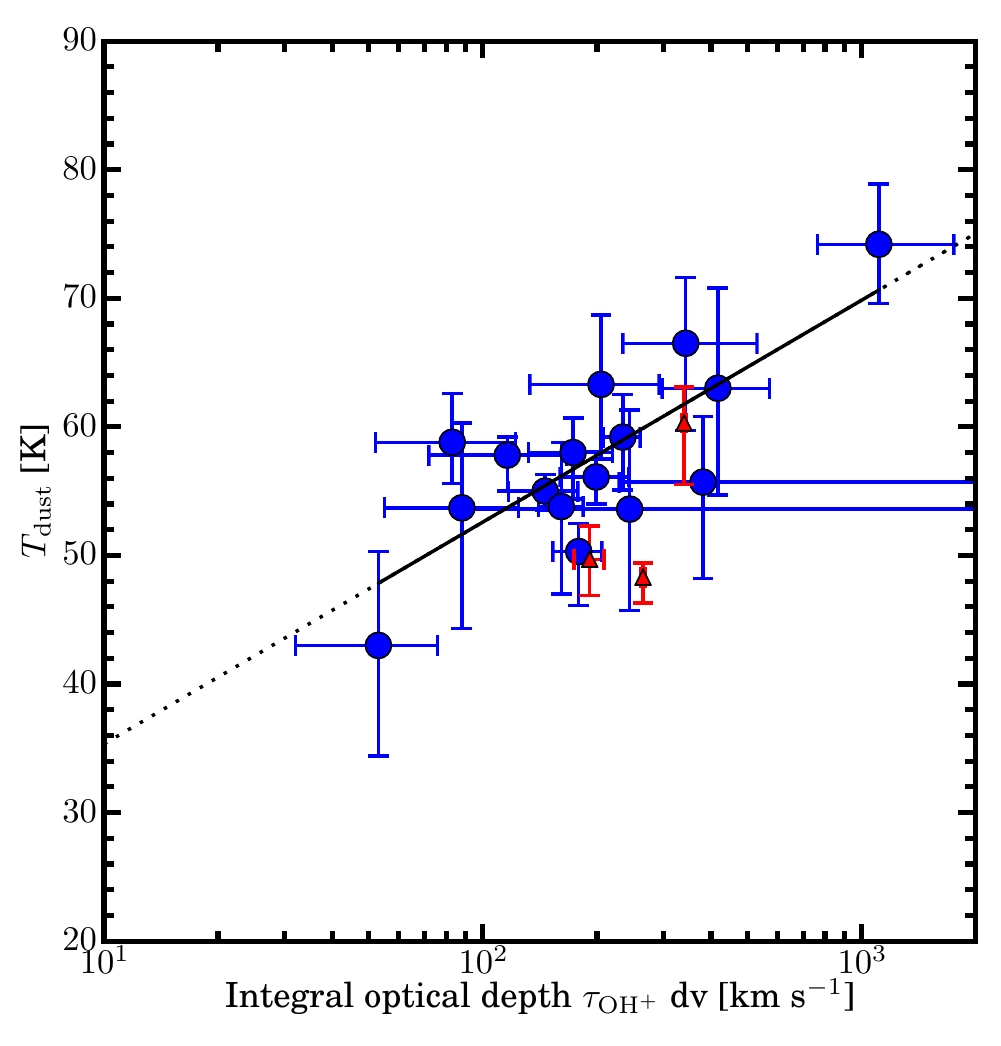}
\plotone{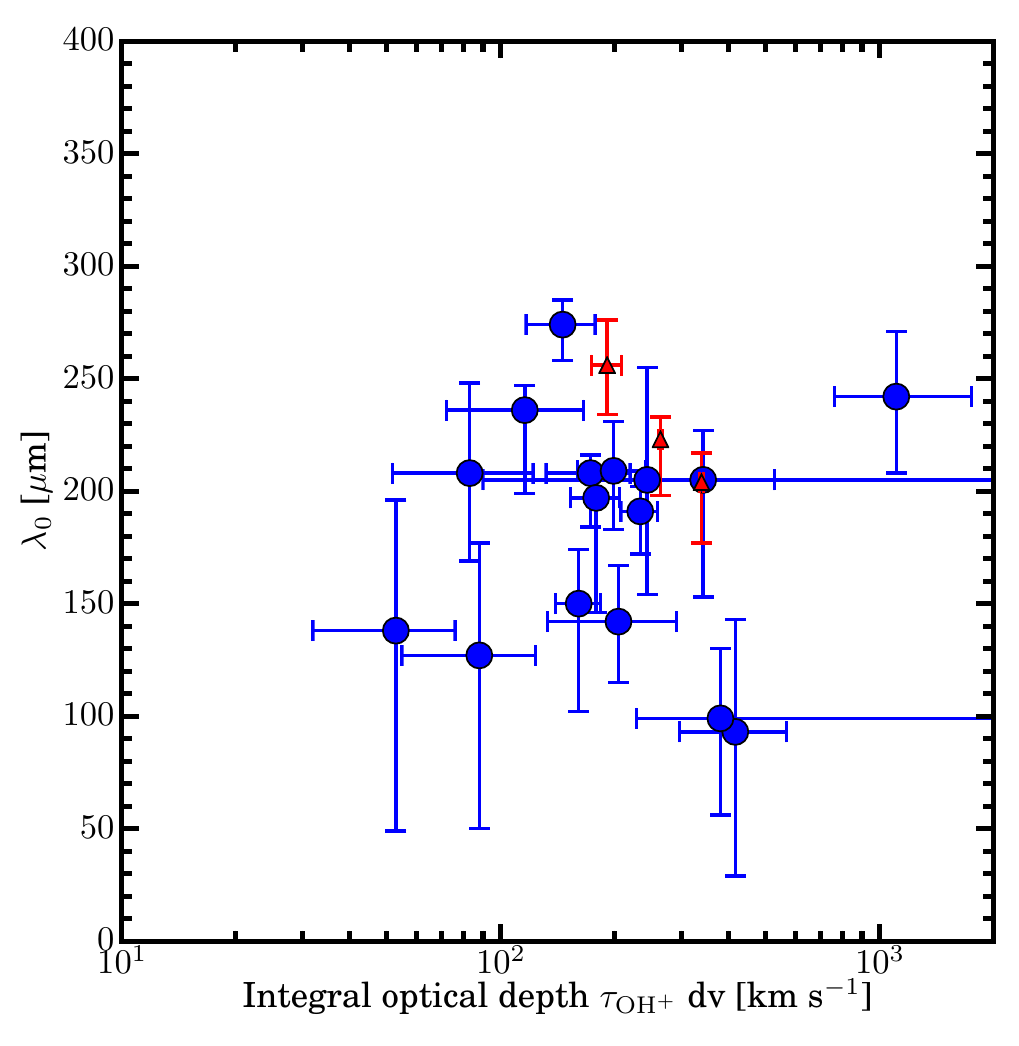}
\plotone{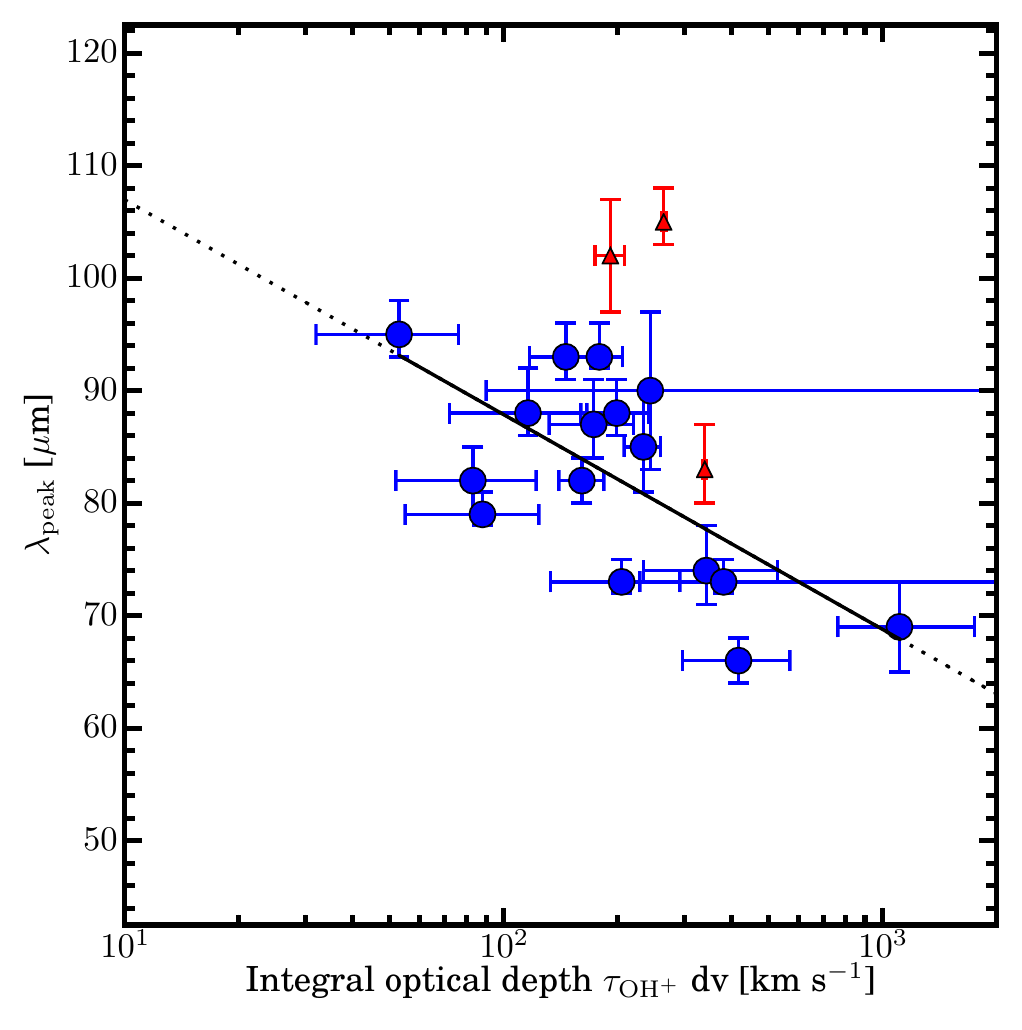}
\vspace{-2mm}

\caption{Relations between the dust temperature, the wavelength where the optical depth reaches unit, and the SED peak wavelength (left to right) as a function of wavelength, OH$^+$ optical depth, and OH$^+$ integral optical depth (top to bottom) for the GADOT sample (blue dots) and three $z$=2--3 DSFGs from the literature (red triangles; see Table~\ref{t3}). For reference, the median values and median absolute deviation of the GADOT sample are shown in the top panels as the dotted green lines and shaded region, and the rest wavelengths of the OH$^+$ transitions in this study are indicated in the top middle panel (showing that the dust optical depth is below unity for the sample at the OH$^+$ wavelengths). There appears to be a correlation (black lines; see Table~\ref{f5} for fit parameters) between the OH$^+$ integral optical depth and the dust temperature (and an inverse correlation with the SED peak wavelength), and a similar but weaker correlation between the OH$^+$ optical depth and the dust temperature (and inversely, with peak wavelength).\label{f11}}
%\vspace{-5mm}
%
\end{figure*}

\begin{figure}
\begin{deluxetable}{ l c c }
%\vspace{-7mm}

\tabletypesize{\scriptsize}
\tablecaption{Best fit dust opacity correlations with a functional form y=$a$+log($b$$\cdot$x/[unit\,x])\,[unit\,y]. \label{t5}}
\tablehead{
Parameters & $a$ & $b$ }
\startdata
$\tau_{\rm OH+}$, $T_{\rm dust}$ & 60.6$\pm$2.0\,K & 5.3$\pm$2.4 \\
$\tau_{\rm OH+}$dv, $T_{\rm dust}$ & 18.1$\pm$8.8\,K & 7.5$\pm$1.6 \\
$\tau_{\rm OH+}$, $\lambda_{\rm peak}$ & 79$\pm$3\,$\mu$m & $-$6.8$\pm$3.1 \\
$\tau_{\rm OH+}$dv, $\lambda_{\rm peak}$ & 126$\pm$13\,$\mu$m & $-$8.3$\pm$2.5 \\
\enddata
\end{deluxetable}
\vspace{-9mm}

\end{figure}

\subsubsection{OH$^+$ Opacity Relations}

The OH$^+$ absorption components are likely associated with cool
($\lesssim$100\,K), low density ($<$100\,cm$^{-3}$),
spatially-extended gas components along the line of sight to the warm
dust continuum emission from the starbursts (e.g., Falgarone et
al.\ \citeyear{falgarone17}; Riechers et al.\ \citeyear{riechers20c}).

To better understand the implications of the OH$^+$ absorption
detections in our sample, we have calculated the peak optical depths
$\tau_{\rm OH+}$=--ln($f_{\rm trans}$) and the velocity-integrated
optical depths $\tau_{\rm OH+}$\,dv (which is a quantity similar to
the line equivalent width) for the GADOT sample. Here, $f_{\rm trans}$
is the fraction of the continuum emission that is still
transmitted. The resulting values are summarized in
Table~\ref{t2b}. The GADOT sample has a median $\tau_{\rm
  OH+}$=0.56$\pm$0.25 and a median $\tau_{\rm
  OH+}$\,dv=220$\pm$110\,\kms. This corresponds to a median OH$^+$
column density of
$N$(OH$^+$)=(1.0$\pm$0.6)$\times$10$^{15}$\,cm$^{-2}$, or a median
hydrogen column density of
$N$(H)=(6.3$\pm$3.8)$\times$10$^{22}$\,cm$^{-2}$ (see Table~\ref{t2b}
for underlying assumptions). For a typical DSFG diameter of 2\,kpc
(e.g., ADFS-27; Riechers et al.\ \citeyear{riechers20c}), this
corresponds to a neutral hydrogen mass of
$M$(H)$\sim$2$\times$10$^9$\,\msol.

We then compare the results to SED-based physical quantities that are,
to first order, independent of lensing magnification. In the top row
of Fig.~\ref{f11}, $T_{\rm dust}$, $\lambda_0$, and $\lambda_{\rm
  peak}$ are shown as a function of redshift. There are no clear
trends except for a potential weak increase in $T_{\rm dust}$ towards
higher redshift, and a corresponding weak blueshifting of
$\lambda_{\rm peak}$. This weak trend is likely primarily due to the
underlying selection function of {\em Herschel} at 250--500\,$\mu$m,
and thus, consistent with no strong underlying redshift evolution
within the uncertainties.

The remaining panels show the same quantities in relation to the peak
and integral optical depths. There is no detectable trend with
$\lambda_0$, such that we do not find evidence for increased OH$^+$
absorption in systems that have higher dust opacity near the
wavelengths of the OH$^+$ lines. This is perhaps expected, because the
OH$^+$ absorption is caused by gas in front of the dust continuum. On
the other hand, there appears to be a significant trend of increased
OH$^+$ absorption depth toward higher dust temperatures (or
correspondingly, towards shorter wavelengths of the SED peak). The
trend appears to be tighter for the integral optical depth
(see Table~\ref{t5}).

There are two potential factors that may explain such a trend. First,
the OH$^+$ 2$\to$1 lines (which reduce the upper level populations of
the 1$_1$$\to$0$_1$ and 1$_2$$\to$0$_1$ transitions if seen in
absorption) lie at $\sim$1892--2029\,GHz, or 148--158\,$\mu$m, i.e.,
close to or shortward of $\lambda_0$ for most of the GADOT
sample. These transitions are observed in absorption in the archetypal
ultra-luminous infrared galaxy (ULIRG) Arp\,220 (Gonzalez-Alfonso et
al.\ \citeyear{ga13}), and tentatively, in the $z$=6.34 starburst
HFLS3 (based on the data shown by Riechers et
al.\ \citeyear{riechers13b}). As such, it is conceivable that the
observed trend could be due to the increased availability of
148--158\,$\mu$m photons as the SED peak shifts toward shorter
wavelengths, which results in an increased pumping efficiency out of
the upper level of the 1$_1$$\to$0$_1$ and 1$_2$$\to$0$_1$
transitions.

Second, the sources with higher $T_{\rm dust}$ may simply be warmer
because they are more compact, such that they have higher $\Sigma_{\rm
  FIR}$ and $\Sigma_{\rm SFR}$ surface densities. Smaller sizes then
lead to a higher density of supernovae once the most massive stars end
their life cycles. Since the supernova rate is proportional to the
rate of cosmic ray production, this would result in an increased
cosmic ray energy density. The central regions of the high $T_{\rm
  dust}$ DSFGs in the GADOT sample thus may resemble giant cosmic ray
dominated regions (CRDRs), rather than ensembles of photon-dominated
regions (PDRs), i.e., similar to what has been discussed for ULIRG
nuclei (e.g., Papadopoulos \citeyear{papadopoulos10}; Gonzalez-Alfonso
et al.\ \citeyear{ga13}). Since cosmic rays are likely the main
production mechanism for molecular ions like OH$^+$, the increased
cosmic ray energy density would then result in an increased abundance
of OH$^+$ for higher $T_{\rm dust}$ sources, causing stronger OH$^+$
absorption and therefore higher OH$^+$ optical depth. High cosmic ray
energy densities could also explain the comparatively high CO
excitation in this sample, as a possible alternative to the
shock-induced CO excitation scenario.

In principle, the OH$^+$ abundance can also be increased in
environments exposed to radiation fields associated with X-ray
dominated regions (XDRs) around AGN (see also discussion by
Gonzalez-Alfonso et al.\ \citeyear{ga13}). At the same time, AGN can
increase the dust temperature as an additional heating source. Such a
contribution can certainly be relevant for local enhancements near an
AGN. However, given the high optical depth of the dust, it is unlikely
that the radiation from AGN can penetrate to large radii even if
buried AGN are present in our sample, such that it remains unclear
that the OH$^+$ production could be sufficiently increased on large
enough scales to fully explain the observed trend.

\section{Conclusions}

We have detected 18 \ico\ emission lines and 23 ground-state OH$^+$
absorption, emission, and P-Cygni-shaped lines toward a sample of 18
{\em Herschel-}selected massive dusty starburst galaxies at
$z$=2.3--6.3 as part of the GADOT Galaxy Survey (including two
previously-published targets). \ico\ emission from warm, dense
molecular gas is detected in all targets, at higher line luminosities
than expected from the $L^\prime_{\rm CO(9-8)}$--$L_{\rm FIR}$
relation for nearby star-forming galaxies. This may suggest that shock
excitation of the molecular gas impacted by galactic winds is more
prevalent in DSFGs in the early universe than in the bulk of nearby
star-forming galaxies, but similar to what is seen in some of the most
active nearby starbursts. The presence of copious amounts of
shock-heated, dense gas is also consistent with the strength of the
OH$^+$ emission in the 2/3 of the sample where it is detected, but
high cosmic ray energy densities would provide a possible alternative
explanation for the observed trends. One of the galaxies is thought to
contain an AGN and shows a spatial offset between the OH$^+$and dust
emission peaks. This may indicate a local enhancement of OH$^+$ due to
the AGN radiation field, but an AGN contribution is not strictly
required to explain its OH$^+$ properties.

OH$^+$ absorption is seen in all but two galaxies in the sample,
showing that it is ubiquitous among high-$z$ starburst galaxies. This
finding on its own appears to imply that the absorbing OH$^+$ gas is
widespread, and not confined to small opening angles, as expected if
it resides in an extended gaseous halo. We find a balance between red-
and blueshifted OH$^+$ absorption/emission components among those
galaxies that show significant velocity shifts relative to CO, which
may suggest that both outflows (as indicated by blueshifted absorption
and/or redshifted emission) and inflows (as indicated by the opposite)
are prevalent. Some of the systems showing evidence for outflows have
outflow velocities that likely exceed the escape velocities of the
galaxies. As such, star formation feedback appears to be important in
the evolution of massive galaxies in their most active phase, but
higher spatial resolution observations are required to study the gas
kinematics and balance of outflows vs. inflows on scales that come
closer to resolving the starburst nuclei for the bulk of the sample
and to better separate systemic gas components from gas flows.

We find a correlation between the OH$^+$ absorption optical depth and
the dust temperature, and an inverse correlation of the former with
the peak wavelength of the dust SED. We suggest that this is due to
the fact that higher $T_{\rm dust}$ DSFGs typically are more compact,
showing higher $\Sigma_{\rm FIR}$ and $\Sigma_{\rm SFR}$. This yields
an increased cosmic ray energy density through a high density of
supernovae, and thus, an increased production of OH$^+$ ions -- hence
stronger OH$^+$ absorption. In principle, this effect could also
contribute to the high CO excitation of the sample. Another
contributing effect may be the enhanced availability of
$\sim$150\,$\mu$m photons for higher $T_{\rm dust}$ DSFGs, which could
lead to an enhanced pumping efficiency out of the upper levels of the
observed OH$^+$ transitions.

Our findings provide an improved understanding of the role of gas
flows in the evolution of massive star-forming galaxies in the early
universe, putting the initial studies of high-$J$ CO and OH$^+$ lines
in distant galaxies on a firmer statistical footing. Detailed studies
show that, when associated with outflows, these gas flows can expel
gas from the star-forming regions at high rates, regulating the growth
of these already massive systems (e.g., Riechers et
al.\ \citeyear{riechers20c}). While inflows were previously thought to
be rare (e.g., Berta et al.\ \citeyear{berta20}), our current
observations may suggest (within a factor of $\sim$1--2.5) a similar
prevalence of OH$^+$ outflows vs.\ inflows, and that it is relatively
common to see OH$^+$ emission components along with the absorption. If
confirmed, this may suggest that the gas inflows are not confined to
small opening angles. On the other hand, the detectability of emission
components from the gas flows in a largely ($>$80\%) strongly
gravitationally lensed galaxy sample with typical magnification
factors of $\sim$5--40 make it likely that the typical sizes of the
entire systems are perhaps only a few kiloparsecs at most, given the
small regions with high amplification factors in most lensing
configurations. This is consistent with the observed sizes of the
continuum emission where magnification factors are known. The high
detection rate suggests that OH$^+$ non-detections, as seen in the
southern component of the (not strongly lensed) $z$=5.66 merger
ADFS-27 (Riechers et al.\ \citeyear{riechers20c}), are relatively rare
among DSFGs, and/or that high spatial resolution observations are
required to identify the galaxy or merger components associated with
the OH$^+$ absorption or emission. It also may suggest that processes
on few kpc scales are key to explain the origin of the OH$^+$
absorption/emission in complex merging systems like ADFS-27, or
similar lower-redshift systems like HXMM-01.

Observations in the short millimeter to submillimeter wavelength
regime with ALMA and NOEMA are critical to give access to a broad
range in line diagnostics in the early universe that {\em Herschel}
has established in the nearby universe based on samples of tens of
galaxies. This study, which was carried out as part of the GADOT
Galaxy Survey, provides the first investigation of a sizeable sample
of high-redshift galaxies in the OH$^+$ diagnostic lines, which
confirms that OH$^+$ is a prime tracer of the physical properties and
chemical composition of the ISM throughout cosmic history, and the
perhaps most easily accessible tracer of molecular outflows and
inflows. Future work on the GADOT sample, together with complementary
investigations (e.g., K.~Butler et al.\ 2021, in preparation), will
expand the standard toolkit for these studies to additional molecular
species, and will provide a more comprehensive study of the CO line
ladders. Since the properties of the ISM set the initial conditions
for star formation, such comprehensive studies are critical to fully
understand the galaxy formation process, and in which ways it may
differ at the high mass, high luminosity end.

\acknowledgments

We thank the anonymous referee for helpful suggestions that led to
some valuable clarifications in this work. The authors thank Daizhong
Liu for sharing the local galaxy source data displayed in the \ico\ --
FIR relation figure. D.R. acknowledges support from the National
Science Foundation under grant numbers AST-1614213 and
AST-1910107. D.R. also acknowledges support from the Alexander von
Humboldt Foundation through a Humboldt Research Fellowship for
Experienced Researchers. A.C. acknowledges support from NASA grant
80NSSC20K0437. The National Radio Astronomy Observatory is a facility
of the National Science Foundation operated under cooperative
agreement by Associated Universities, Inc. This paper makes use of the
following ALMA data: ADS/JAO.ALMA\#\,2016.2.00105.S; 2018.1.00922.S;
2017.1.00235.S; and 2018.1.00966.S. ALMA is a partnership of ESO
(representing its member states), NSF (USA) and NINS (Japan), together
with NRC (Canada) and NSC and ASIAA (Taiwan), in cooperation with the
Republic of Chile.

\vspace{5mm}
\facilities{ALMA, ATCA, Hubble(ACS and WFC3)}

\software{CASA package (v5.6.1; McMullin et al.\ \citeyear{mcmullin07})}

\bibliographystyle{yahapj}
\bibliography{ref.bib}

\end{document}